\RequirePackage{fix-cm}
\documentclass[smallextended]{svjour3}       
\smartqed  
\usepackage{graphicx}
\usepackage{natbib}
\setcitestyle{aysep={}}
\newcommand\cites[1]{\citeauthor*{#1}'s\ (\citeyear{#1})}	
\usepackage{xcolor}		
\usepackage{amssymb}	
\usepackage{amsmath}	
\usepackage{multirow}
\usepackage{tabularx}
\usepackage{enumitem} 	
\usepackage{pdflscape}
\usepackage{afterpage}
\begin{document}

\title{Specific investments under negotiated transfer pricing: effects of different surplus sharing parameters on managerial performance
}
\subtitle{An agent-based simulation with fuzzy Q-learning agents}

\titlerunning{Specific investments under negotiated transfer pricing}        

\author{Christian Mitsch
}


\institute{Christian Mitsch (0000-0002-8587-4887) \at
              Department of Management Control and Strategic Management, University of Klagenfurt \\
              Universit\"atsstra\ss{}e 65-67, 9020 Klagenfurt, Austria \\
              \email{christian.mitsch@aau.at}           
}

\date{}


\vspace*{3mm}
\begin{center}
 	\begin{Large}
 		\textbf{Specific investments under negotiated transfer pricing: effects of different surplus sharing parameters on managerial performance} \\
 	\end{Large}
 	\vspace{3mm}
 	\begin{large}
 		An agent-based simulation with fuzzy Q-learning agents \\
 	\end{large}
 	\vspace{10mm}
 	\begin{normalsize}
 		$\text{Christian Mitsch}^{[0000-0002-8587-4887]}$ \\
 		\vspace{3mm}
 		Department of Management Control and Strategic Management \\
 		University of Klagenfurt, 9020, Austria \\
 		\texttt{christian.mitsch@aau.at} 
              
 	\end{normalsize}
\end{center}

\vspace{10mm}

\begin{abstract}

	This paper focuses on a decentralized profit-center firm that uses negotiated transfer pricing as an instrument to coordinate the production process. 
	Moreover, the firm's headquarters gives its divisions full authority over operating decisions and it is assumed that each division can additionally make an upfront investment decision that enhances the value of internal trade. 
	On early works, the paper expands the number of divisions by one downstream division and relaxes basic assumptions, such as the assumption of common knowledge of rationality.
	Based on an agent-based simulation, it is examined whether cognitively bounded individuals modeled by fuzzy Q-learning achieve the same results as fully rational utility maximizers. 
	In addition, the paper investigates different constellations of bargaining power to see whether a deviation from the recommended optimal bargaining power leads to a higher managerial performance. 
	The simulation results show that fuzzy Q-learning agents perform at least as well or better than fully individual rational utility maximizers. 
	The study also indicates that, in scenarios with different marginal costs of divisions, a deviation from the recommended optimal distribution ratio of the bargaining power of divisions can lead to higher investment levels and, thus, to an increase in the headquarters' profit. 

\vspace{5mm}

\keywords{Agent-based simulation \and Fuzzy Q-learning \and Hold-up problem \and Negotiated transfer pricing \and Specific investments}
\end{abstract}

\section{Introduction}
\label{sec:Introduction}


		In decentralized firms, the headquarters gives its divisions more authority in making decisions. 
		Transfer pricing and managerial performance evaluation are two key instruments for managing potential conflicts between divisions and the headquarters and guiding the internal trade between divisions \citep[e.g.,][]{anctil1999negotiated, baldenius1999negotiated}. 
		Internal transfers, especially in multi-stage production processes, are often performed under conditions of asymmetric information \citep{baldenius2000intrafirm}. 
		To induce effort and reduce the risk of moral hazard, divisional profit-based compensation schemes are often applied \citep[e.g.,][]{edlin1995specific, vaysman1998model}. 
		
		The divisions' coordination problem increases when divisions can additionally make specific investments that enhance the value of internal trade. 
		The main problem with specific investments is that they are irreversible and are of little or no value in the divisions' external lines of business \citep{edlin1995specific}. 	
		Since those investments are usually made under uncertainty, each division tends to underinvest. 
		This problem is also known as an investment ``hold-up'' problem \citep{schmalenbach1908verrechnungspreise, williamson1979transaction, williamson1985economic}. 
	
		While various methods of transfer pricing to mitigate the hold-up problem have been investigated extensively in the economic transfer pricing literature \citep[e.g.,][]{baldenius1999negotiated, edlin1995specific, hofmann2006verfugungsrechte, lengsfeld2004transfer, pfeiffer2011cost, wagner2008corporate}, it seems less clear how well the literature's recommendations actually work in practice.  
		Moreover, the solutions of transfer pricing problems are often based on game theory approaches, such as the concept of subgame perfect equilibrium. 
		Such concepts require demanding assumptions, e.g., of rational decision-making behavior or common knowledge, and, in practice, these assumptions are often not met \citep{axtell2007economic, simon1979rational}. 
		Assumptions on rationality are questionable in models that reflect human behavior and, therefore, most of such models quickly lose practical relevance \citep{young2001individual}. 
	
		Against this background, this paper focuses on negotiated transfer pricing and, specifically, addresses and extends the simulation model introduced in \cite{mitsch2023impact}. 
		In particular, the simulation study analyzes a decentralized firm in which divisions produce highly specialized intermediate products. 
		The firm's headquarters applies the concept of profit centers to determine the profit for internal divisions of responsibility. 
		In addition, divisions can make upfront specific investments and have private information regarding their area of responsibility. 
		A linear divisional profit-based compensation scheme is applied in order to guarantee that the divisions act in the headquarters' interest. 
		Furthermore, the study relaxes basic assumptions, e.g., the assumption of common knowledge of rationality for the divisions, and examines whether cognitively bounded individuals (modeled with fuzzy Q-learning agents) achieve the same results as fully rational utility maximizers. 
		In addition, different constellations of bargaining power are examined to see whether a deviation from the recommended optimal bargaining power leads to a higher divisional performance. 
		The findings of \cites{mitsch2023impact} simulation study show that cognitively bounded individuals can achieve the same results as fully rational utility maximizers, but, in certain cases where divisional investment costs differ widely from each other, cognitively bounded individuals can generate even higher profits, if they obtain approximately the same level of bargaining power from the headquarters for their negotiation process. 
		
		In this study, the number of divisions is increased by one. 
		Concretely, there is one supplying division and two buying divisions. 
		As a consequence, two negotiations over transfer prices and quantities have to be carried out at the same time. 
		Since the upfront investment decisions depend on the anticipated outcome of the negotiations, it is more difficult for the headquarters to determine the optimal bargaining power for each division. 
		Simultaneously, the investments made by the divisions cannot be individually separated out due to their nature. 
		Hence, the management decision problem to be solved cannot be divided into two or more disjoint parts. 
		
		This paper distinguishes between an ``all-knowing'' headquarters (serve as a benchmark for optimal decision-making behavior) and a headquarters that does not have all the information to find the optimal bargaining power for its divisions and, therefore, relies on reference values or simple rules to determine the bargaining power of each division. 
		The first research question in this contribution is to examine whether cognitively bounded individuals achieve the same results as fully rational utility maximizers on the investment hold-up problem described above.
		In addition, the study also seeks to examine the impact of bargaining power on managerial performance. 
		
		For this purpose, this paper conducts an agent-based computer simulation with individual learning agents which are modeled by fuzzy Q-learning. 
		The use of fuzzy Q-learning offers many advantages, including a high degree of heterogeneity with respect to the structure and the dynamic interactions between agents and their environment and, in particular, it is a feasible way to deal with the divisions' bounded rationality and cognitive limitations. 
		Moreover, an agent-based simulation also allows to observe the agents' behavior as well as the system's behavior on the macro-level in time, which otherwise cannot be derived in relation to a ``functional relationship'' from the individual behaviors of those agents \citep[e.g.,][]{epstein2006generative, wall2016agent}. 

		The remainder of this paper is organized as follows: In Sec. \ref{sec:The_Model}, the extended negotiated transfer pricing model is introduced and the resulting solutions are discussed. 
		Section \ref{sec:agent_based_simulation} introduces the simulation model with fuzzy Q-learning agents. 
		The parameter settings for the agent-based simulation are explained in Sec. \ref{sec:Parameter_Settings_and_Simulation_Setup} and, in Sec.  \ref{sec:Results_and_Discussion}, the simulation results are presented and discussed. 
		Section \ref{sec:Conclusion} contains concluding remarks and suggests possible directions for future research. 

\newpage
\section{Negotiated transfer pricing model}
\label{sec:The_Model}

\subsection{Specification of the firm} 
\label{ssec:Organizational_Setting}

\begin{figure}[h!]
		\centering
		\includegraphics[width=0.9\textwidth]{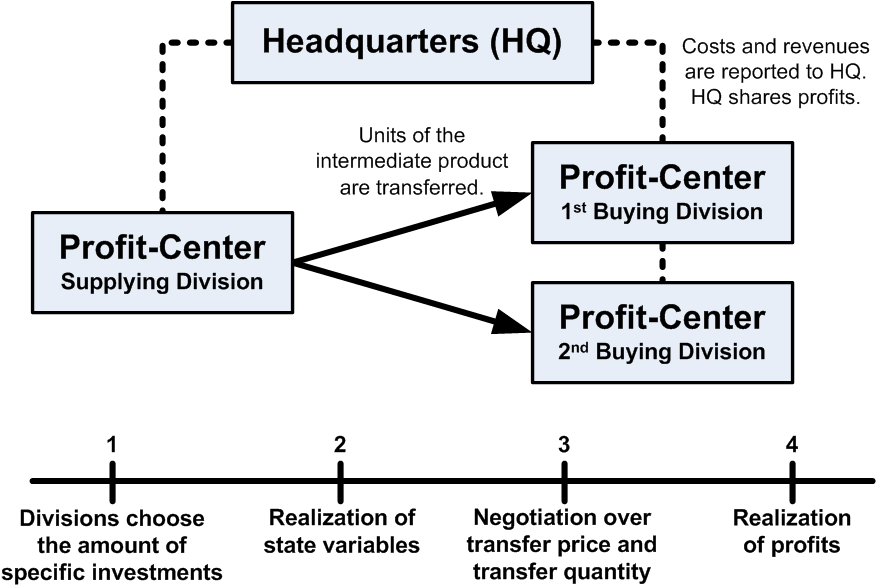}
		\caption{Schematic representation of the decentralized firm. In addition, the sequence of events within the negotiated transfer pricing model is illustrated. Source: Based on \cite{wagner2008corporate} and, further, modified here for the case of two buying divisions.}
		\label{fig:Schematic_Representation}
	\end{figure}

	In this paper, a decentralized firm is examined that consists of one headquarters and three divisions - one supplying division (the ``upstream'' division) and two buying divisions (the ``downstream'' divisions). 
	Suppose that the supplying division purchases raw materials from a commodity market in order to produce a highly specialized intermediate product that, in particular, cannot be bought from an external market \citep[e.g.,][]{anctil1999negotiated, edlin1995specific, wagner2008corporate}.
	On the downside, each buying division refines the intermediate products independently of each other in order to improve their quality.
	Lastly, the buying divisions sell the refined products on an outlet market separately. 
	Beyond that, all three divisions can additionally make specific investments that increase the value of internal trade. 
	However, these investment decisions have to be made before the negotiations over transfer prices and transfer quantities take place. 
	Figure \ref{fig:Schematic_Representation} schematically represents the negotiated transfer pricing model investigated here. 
	
	Corresponding to the well-known negotiated transfer pricing models by \cite{eccles1988price}, \cite{edlin1995specific}, \cite{gox2006economic}, \cite{pfeiffer2007rekonstruktion}, \cite{vaysman1998model}, and \cite{wagner2008corporate}, it is assumed that all three divisions are treated as profit centers. 
	Therefore, divisional performance evaluations are based on profits and each division is allowed to determine the amount of specific investments, the level of the transfer price, and the amount of intermediate products autonomously. 
	
	\vspace{0mm}

	In the following, the supplying division, the $1$st buying division, the $2$nd buying division, and the headquarters are abbreviated to $S$, $1$, $2$, and $HQ$, respectively. 
	To keep the analysis simple and, especially to ensure, that the negotiated transfer pricing model has a unique subgame perfect equilibrium, it is assumed that the supplying division's costs of manufacturing $q_j \in \mathbb{R}^+$, $j \in \{ 1,2 \}$, units of the intermediate product are given by 
	\begin{equation}
		C_S(q_1,q_2,\theta_S,I_S) = (\theta_S - I_S ) \cdot (q_1 + q_2) \;,  \label{eq:supplying_division_costs}	
	\end{equation}
	where $\theta_S \in \mathbb{R}^+$ is a state variable which reflects the purchase price of raw materials and $I_S \in \mathbb{R}^+$ stands for the amount of specific investment carried by the supplying division. 
	In contrast, each buying division's, $j \in \{ 1,2 \}$, net revenue is 
	\begin{equation}
		R_j(q_j,\theta_j,I_j) = (\theta_j - \frac{1}{2} \; b \; q_j + I_j) \; q_j \; , \label{eq:buying_division_net_revenue}
	\end{equation}  
	where $\theta_j \in \mathbb{R}^+$ is a state variable which represents the constant term in the inverse demand function for the selling product and $I_j \in \mathbb{R}^+$ denotes the amount of specific investment carried by the $j$th buying division. 
	For the sake of simplicity, it is assumed that $b_1 = b_2 = b \in \mathbb{R}^+$, where $b$ describes the slope of the inverse demand function.
	
	In the decentralized setting examined here, it is assumed that the supplying division has private information about purchase prices of raw materials on the commodity market, while the buying divisions have private information about selling prices of the refined products on the outlet market. 
	Therefore, the supplying division and the $j$th buying division know the distribution of the state variable $\theta_S$ and $\theta_j$, respectively.
	In order to solve the transfer price problem analytically, it is required that all divisions have at least access to information about the expected values of the markets.
	For the sake of simplicity, it is assumed that the state variables $\theta_S$, $\theta_1$, and $\theta_2$ are stochastically independent random variables. 
	Apart from that, the analysis assumes that the headquarters cannot observe the state variables nor the undertaken investments; 
	the headquarters only knows the expected values of the state variables.
	Furthermore, the headquarters' accounting system receives costs and revenues after the negotiation phase and, subsequently, the headquarters' profit is calculated by 
	\begin{equation}
		\Pi_{HQ}(q_1,q_2,\theta_S,\theta_1,\theta_2,I_S,I_1,I_2) = R_1 + R_2 - C_S - w_S - w_1 - w_2 \; .  \label{eq:headquarters_profit}
	\end{equation}
	The supplying division's investments as well as each buying division's investments cause divisional investment costs (or divisional capital expenditure) $w_S(I_S)$ and $w_j(I_j)$, respectively.
	As in the study of \cite{baldenius1999negotiated}, \cite{edlin1995specific}, \cite{hofmann2006verfugungsrechte}, \cite{pfeiffer2007rekonstruktion}, and \cite{wagner2008corporate}, the divisional investment costs have the following quadratic cost structure. 
	\begin{equation}
		w_j(I_j) = \frac{1}{2} \; \lambda_j \; I_j^2 \; \; \; \text{for } j \in \{ S,1,2\} \label{eq:divisional_investment_costs}
	\end{equation}
	The marginal cost parameter is denoted by $\lambda_j \in \mathbb{R}^+$ 
	and the analysis assumes that the parameters $w_S$, $w_1$, $w_2$, $\lambda_S$, $\lambda_1$, $\lambda_2$, and $b$ are known to the headquarters and each division. 
	The term, one half, is only for reasons of expediency. 
	
	The sequence of events within the negotiated transfer pricing model (see Fig. \ref{fig:Schematic_Representation}) can be summarized as follows: 
	at date one, all three divisions have to make an investment decision independently of each other. 
	Subsequently, the supplying division and each buying division independently observe the state variables $\theta_S$ and $\theta_j$, respectively. 
	At date three, the amount of intermediate products is determined by the divisions and, finally, profits are realized. 

	\subsection{First-best solution}
	\label{ss:First_Best_Solution}

		In order to provide a benchmark solution (first-best solution or ex post efficient solution), Eq. \ref{eq:headquarters_profit} is maximized with respect to investment and quantity. 
		Since investment and quantity are determined at different times, the analysis starts by backward induction on date three, on which the quantity is set.\footnote{ 
		A step-by-step solution of the two-stage decision problem for $j=1$ (only one buying division) is presented, e.g., in \cites{wagner2008corporate} PhD thesis.} 
		\begin{equation}
			(q_1^*,q_2^*) \in \underset{(q_1,q_2) \in \mathbb{R}_+^2}{\mathrm{arg\,max}} \; \Pi_{HQ}(q_1,q_2,\theta_S,\theta_1,\theta_2,I_S,I_1,I_2) 
		\end{equation}
		In this paper, first-best solutions are indexed by a superscript $^*$. 
		On date three, the first order condition for maximizing the headquarters' profit $\Pi_{HQ}$ with respect to $q_j$ is equal to 
		\begin{equation}
			\frac{\partial \Pi_{HQ}}{\partial q_j} = \theta_j - b \; q_j + I_j - \theta_S + I_S = 0 
		\end{equation}
		and, hence, the profit maximizing quantity is given by
		\begin{equation}
			q_j^* = \frac{\theta_j - \theta_S + I_S + I_j}{b} \; , \text{ for } j \in \{ 1,2 \} \; . \label{eq:backward_q}
		\end{equation}
		On date one, the investment decisions are made under uncertainty and, therefore, the expected headquarters' profit $E[\Pi_{HQ}]$ is maximized with respect to $I_S$, $I_1$, and $I_2$. 
		\begin{equation}
			(I_S^*,I_1^*,I_2^*) \in \underset{(I_S,I_1,I_2) \in \mathbb{R}_+^3}{\mathrm{arg\,max}} \; E[\Pi_{HQ}(q_1,q_2,\theta_S,\theta_1,\theta_2,I_S,I_1,I_2)] 	\label{eq:backward_I}
		\end{equation}
		Differentiating the expected headquarters' profit with respect to investments yields 
		\begin{eqnarray}
			I_S^* & = & \frac{E[q_1^*+q_2^*]}{\lambda_S} \; , \label{eq:backward_I_S} \\
			I_j^* & = & \frac{E[q_j^*]}{\lambda_j} \; , \text{ for } j \in \{ 1,2 \} \; . \label{eq:backward_I_j} 
		\end{eqnarray}
		Now, substituting Eq. \ref{eq:backward_q} into Eq. \ref{eq:backward_I_S} and \ref{eq:backward_I_j}, one gets the following first-best expected investments. 
		\begin{eqnarray}
			I_S^* & = & \frac{E[\theta_1 - \theta_S] \cdot (b \; \lambda_1 \; \lambda_2 - \lambda_1) + E[\theta_2 - \theta_S] \cdot (b \; \lambda_1 \; \lambda_2 - \lambda_2)}{b^2 \; \lambda_1 \; \lambda_2 \; \lambda_S - b \; ( \lambda_1 \; \lambda_S + \lambda_2 \; \lambda_S +2 \; \lambda_1 \; \lambda_2) + \lambda_1 + \lambda_2 + \lambda_S} \label{eq:optimal_I_S} \\
			I_1^* & = & \frac{E[\theta_1 - \theta_S] \cdot (b \; \lambda_2 \; \lambda_S - \lambda_2 - \lambda_S) + E[\theta_2 - \theta_S] \cdot \lambda_2}{b^2 \; \lambda_1 \; \lambda_2 \; \lambda_S - b \; ( \lambda_1 \; \lambda_S + \lambda_2 \; \lambda_S +2 \; \lambda_1 \; \lambda_2) + \lambda_1 + \lambda_2 + \lambda_S} \label{eq:optimal_I_1} \\
			I_2^* & = & \frac{E[\theta_2 - \theta_S] \cdot (b \; \lambda_1 \; \lambda_S - \lambda_1 - \lambda_S) + E[\theta_1 - \theta_S] \cdot \lambda_1}{b^2 \; \lambda_1 \; \lambda_2 \; \lambda_S - b \; ( \lambda_1 \; \lambda_S + \lambda_2 \; \lambda_S +2 \; \lambda_1 \; \lambda_2) + \lambda_1 + \lambda_2 + \lambda_S} \label{eq:optimal_I_2}
		\end{eqnarray}
		Finally, substituting Eq. \ref{eq:optimal_I_S} - \ref{eq:optimal_I_2} into Eq. \ref{eq:backward_q} implies the first-best expected quantities. 
		\begin{small}
		\begin{eqnarray}
			q_1^* & = & \frac{\theta_1-\theta_S}{b} + \frac{E[\theta_1-\theta_S] (b \lambda_1 \lambda_2 + b \lambda_2 \lambda_S - \lambda_1 - \lambda_2 - \lambda_S) + E[\theta_2-\theta_S] \; b \lambda_1 \lambda_2}{b (b^2 \; \lambda_1 \; \lambda_2 \; \lambda_S - b (\lambda_1 \; \lambda_S + \lambda_2 \; \lambda_S +2 \; \lambda_1 \; \lambda_2) + \lambda_1 + \lambda_2 + \lambda_S)} \label{eq:optimal_q_1} \\
			q_2^* & = & \frac{\theta_2-\theta_S}{b} + \frac{E[\theta_2-\theta_S] (b \lambda_1 \lambda_2 + b \lambda_1 \lambda_S - \lambda_1 - \lambda_2 - \lambda_S) + E[\theta_1-\theta_S] \; b \lambda_1 \lambda_2}{b (b^2 \; \lambda_1 \; \lambda_2 \; \lambda_S - b (\lambda_1 \; \lambda_S + \lambda_2 \; \lambda_S +2 \; \lambda_1 \; \lambda_2) + \lambda_1 + \lambda_2 + \lambda_S)} \label{eq:optimal_q_2}
		\end{eqnarray}
		\end{small}
		\hspace{-2.5mm} With the first-best solutions of the two-stage decision problem, the first-best expected profit is given by the following expression. 
		\begin{small}
		\begin{equation}
			\Pi_{HQ}^* \hspace{-0.5mm}=\hspace{-0.7mm} \frac{E[\theta_1 \hspace{-1mm}-\hspace{-0.7mm} \theta_S]^2 (b \; \lambda_1 \lambda_2 \lambda_S \hspace{-1mm}-\hspace{-0.7mm} \lambda_1 \lambda_S) \hspace{-0.5mm}+\hspace{-0.5mm} E[\theta_2\hspace{-1mm}-\hspace{-0.7mm}\theta_S]^2 (b \; \lambda_1 \lambda_2 \lambda_S \hspace{-1mm}-\hspace{-0.7mm} \lambda_2 \lambda_S) \hspace{-1mm}-\hspace{-0.7mm} E[\theta_1\hspace{-1mm}-\hspace{-0.7mm}\theta_2]^2 \lambda_1 \lambda_2}{2 \; (b^2 \; \lambda_1 \; \lambda_2 \; \lambda_S - b (\lambda_1 \; \lambda_S + \lambda_2 \; \lambda_S +2 \; \lambda_1 \; \lambda_2) + \lambda_1 + \lambda_2 + \lambda_S)} \label{eq:optimal_Profit_HQ}
		\end{equation}
		\end{small}
		\hspace{-2.5mm} The headquarters' profit in Eq. \ref{eq:optimal_Profit_HQ} can be seen as a benchmark for the highest feasible profit that can be achieved, if investments and quantities are set according to Eq. \ref{eq:optimal_I_S} - \ref{eq:optimal_I_2} and Eq. \ref{eq:optimal_q_1} - \ref{eq:optimal_q_2}, respectively. 

	\subsection{Second-best solution}
	\label{sec:Second_Best_Solution}

		Assuming that each division acts in its own interest, the first-best solutions are usually not achieved. 
		However, the headquarters could provide the following linear divisional profit-based compensation schemes to ensure that the negotiations over transfer prices and quantities result in ex post efficient transfer prices and quantities.\footnote{ 
		In general, incentives for efficient investments do not necessarily provide incentives for an efficient quantity and verse versa. For instance, in full-cost transfer pricing models, both divisions are rewarded for efficient investments, but the negotiation does not lead to an ex post efficient quantity \citep[][]{baldenius1999negotiated}.} 
		\begin{eqnarray}
			\Pi_S(q_1,q_2,\theta_S,\theta_1,\theta_2,I_S,I_1,I_2,\Gamma_1,\Gamma_2) & = & \Gamma_1 \; M_1 + \Gamma_2 \; M_2 - w_S \label{eq:Profit_S} \\
			\Pi_1(q_1,\theta_S,\theta_1,I_S,I_1,\Gamma_1) & = & (1-\Gamma_1) \; M_1 - w_1 \label{eq:Profit_1} \\
			\Pi_2(q_2,\theta_S,\theta_2,I_S,I_2,\Gamma_2) & = & (1-\Gamma_2) \; M_2 - w_2 \label{eq:Profit_2} 
		\end{eqnarray}	
		$\Pi_S$, $\Pi_1$, and $\Pi_2$ denote the profit of the supplying division, the profit of the $1$st buying division, and the profit of the $2$nd buying division, respectively. 
		$M_j$ stands for the headquarters' contribution margin with regard to the internal trade between supplying division and the $j$th buying division, i.e., $M_j=(\theta_j -\frac{1}{2} \hspace{0.5mm} b \hspace{0.5mm} q_j + I_j) \hspace{0.5mm} q_j - (\theta_S-I_S) \hspace{0.5mm} q_j$, $j \in \{ 1,2 \}$. 
		$\Gamma_j \in [0,1]$ represents the share of the contribution margin achieved between the supplying division and the $j$th buying division (also known as the $\Gamma$-surplus sharing rule \citep{edlin1995specific}, the supplying division's bargaining power \citep{baldenius1999negotiated}, or the surplus sharing parameter \citep{wielenberg2000negotiated}). 

		As in the preceding section, the headquarters can apply the concept of subgame perfect equilibrium. 
		By doing so, the analysis of negotiated transfer pricing starts by backward induction on date three. 
		Starting with the quantity decision on date three, one gets
		\begin{equation}
			q_j^{sb} = \frac{\theta_j - \theta_S + I_S + I_j}{b} \; , \text{ for } j \in \{ 1,2 \} \; . \label{eq:backward_q_sb}
		\end{equation}
		Note that differentiating the divisional profits with respect to $q_j$, $j \in \{ 1,2 \}$, leads to the same profit maximizing quantity given by Eq. \ref{eq:backward_q}, but the first order conditions for maximizing the expected divisional profits $E[\Pi_j]$ with respect to $I_j$, $j \in \{ S,1,2 \}$, lead to second-best investments (labelled by the superscript $sb$). 
		\begin{eqnarray}
			I_S^{sb} & = & \frac{\Gamma_1 \; E[q_1^{sb}] + \Gamma_2 \; E[q_2^{sb}]}{\lambda_S} \label{eq:second_best_I_j} \\
			I_j^{sb} & = & \frac{(1-\Gamma_j) \; E[q_j^{sb}]}{\lambda_j} \; , \text{ for } j \in \{ 1,2 \}
		\end{eqnarray}
		Since the headquarters has to specify how the contribution margins are shared among the three divisions, the headquarters has to solve the following additional decision problem before the decision-making process for the divisions takes place. 
		\begin{equation}
			(\Gamma_1^{sb},\Gamma_2^{sb}) \in \underset{(\Gamma_1,\Gamma_2) \in [0,1]^2}{\mathrm{arg\,max}} \; E[\Pi_{HQ}(q_1,q_2,\theta_S,\theta_1,\theta_2,I_S,I_1,I_2,\Gamma_1,\Gamma_2)] 
		\end{equation}

		\afterpage{
		\begin{landscape}
		\hspace{-5.3mm} Differentiating the expected headquarters' profit with respect to the surplus sharing parameter $\Gamma_j$, for $j \in \{ 1,2 \}$, 
		\begin{small}
		\begin{equation}
			\frac{\partial}{\partial \Gamma_j} \Big( \big( E[\theta_1]-\frac{1}{2} b q_1^{sb} + I_1^{sb} \big) q_1^{sb} - \big( E[\theta_S] - I_S^{sb} \big) q_1^{sb}  + \big( E[\theta_2]-\frac{1}{2} b q_2^{sb} + I_2^{sb} \big) q_2^{sb} - \big( E[\theta_S] - I_S^{sb} \big) q_2^{sb}  -  \frac{1}{2} \lambda_S {I_S^{sb}}^2 - \frac{1}{2} \lambda_1 {I_1^{sb}}^2 - \frac{1}{2} \lambda_2 {I_2^{sb}}^2 \Big)
		\end{equation}
		\end{small}
		and setting that equation to zero, yields to the following two expressions.
		\begin{small}		
		\begin{eqnarray}
			\Gamma_1^{sb} = \frac{E[\theta_1\hspace{-1mm}-\hspace{-0.7mm}\theta_S] \lambda_S(b \lambda_1 (\lambda_1+\lambda_2-b \lambda_1 \lambda_2) - \lambda_1)   +   E[\theta_2\hspace{-1mm}-\hspace{-0.7mm}\theta_S] \lambda_S(b \lambda_1 \lambda_2 (2-b \lambda_1)-\lambda_2)}{E[\theta_1\hspace{-1mm}-\hspace{-0.7mm}\theta_S]\lambda_1 \lambda_S (b(\lambda_1\hspace{-1mm}+\hspace{-0.7mm}4\lambda_2\hspace{-1mm}+\hspace{-0.7mm}\lambda_S) \hspace{-1mm}-\hspace{-0.7mm} b^2 \lambda_2(\lambda_1\hspace{-1mm}+\hspace{-0.7mm}\lambda_2\hspace{-1mm}+\hspace{-0.7mm}\lambda_S)\hspace{-1mm}-\hspace{-0.7mm}3)   +   E[\theta_2\hspace{-1mm}-\hspace{-0.7mm}\theta_S] \lambda_2 \lambda_S(b \lambda_1\hspace{-1mm}-\hspace{-0.7mm}1)   +   E[\theta_1\hspace{-1mm}-\hspace{-0.7mm}\theta_2] \lambda_1 \lambda_2(b \lambda_1 \hspace{-1mm}+\hspace{-0.7mm} b \lambda_2 \hspace{-1mm}-\hspace{-0.7mm}2)} \label{eq:Gamma_1_sb} \\
			\Gamma_2^{sb} = \frac{E[\theta_2\hspace{-1mm}-\hspace{-0.7mm}\theta_S] \lambda_S(b \lambda_2 (\lambda_1+\lambda_2-b \lambda_1 \lambda_2) - \lambda_2)   +   E[\theta_1\hspace{-1mm}-\hspace{-0.7mm}\theta_S] \lambda_S(b \lambda_1 \lambda_2 (2-b \lambda_2)-\lambda_1)}{E[\theta_2\hspace{-1mm}-\hspace{-0.7mm}\theta_S]\lambda_2 \lambda_S (b(\lambda_2\hspace{-1mm}+\hspace{-0.7mm}4\lambda_1\hspace{-1mm}+\hspace{-0.7mm}\lambda_S) \hspace{-1mm}-\hspace{-0.7mm} b^2 \lambda_1(\lambda_1\hspace{-1mm}+\hspace{-0.7mm}\lambda_2\hspace{-1mm}+\hspace{-0.7mm}\lambda_S)\hspace{-1mm}-\hspace{-0.7mm}3)   +   E[\theta_1\hspace{-1mm}-\hspace{-0.7mm}\theta_S] \lambda_1 \lambda_S(b \lambda_2\hspace{-1mm}-\hspace{-0.7mm}1)   +   E[\theta_2\hspace{-1mm}-\hspace{-0.7mm}\theta_1] \lambda_1 \lambda_2(b \lambda_1 \hspace{-1mm}+\hspace{-0.7mm} b \lambda_2 \hspace{-1mm}-\hspace{-0.7mm}2)} \label{eq:Gamma_2_sb}
		\end{eqnarray}
		\end{small}
		By using substitution as in Sec. \ref{ss:First_Best_Solution}, one obtains the following second-best solutions.  
		\begin{small}		
		\begin{eqnarray}
			I_S^{sb} & = & \frac{(\Gamma_1 E[\theta_1-\theta_S] + \Gamma_2 E[\theta_2-\theta_S]) b \lambda_1 \lambda_2   -   E[\theta_1-\theta_S] \Gamma_1 (1-\Gamma_2) \lambda_1   -   E[\theta_2-\theta_S] (1-\Gamma_1)\Gamma_2 \lambda_2}{b^2 \lambda_1 \lambda_2 \lambda_S - b ((1\hspace{-1mm}-\hspace{-0.7mm}\Gamma_1) \lambda_2 \lambda_S + (1\hspace{-1mm}-\hspace{-0.7mm}\Gamma_2) \lambda_1 \lambda_S + (\Gamma_1\hspace{-1mm}+\hspace{-0.7mm}\Gamma_2) \lambda_1 \lambda_2)   +   (1\hspace{-1mm}-\hspace{-0.7mm}\Gamma_1)(1\hspace{-1mm}-\hspace{-0.7mm}\Gamma_2) \lambda_S + \Gamma_1(1\hspace{-1mm}-\hspace{-0.7mm}\Gamma_2) \lambda_1   +   (1\hspace{-1mm}-\hspace{-0.7mm}\Gamma_1)\Gamma_2 \lambda_2} \label{eq:second_best_I_S} \\
			I_1^{sb} & = & \frac{E[\theta_1-\theta_S] (1-\Gamma_1) b \lambda_2 \lambda_S   -   E[\theta_1-\theta_S](1-\Gamma_1)(1-\Gamma_2) \lambda_S   -   E[\theta_1-\theta_2] (1-\Gamma_1)\Gamma_2 \lambda_2}{b^2 \lambda_1 \lambda_2 \lambda_S - b ((1\hspace{-1mm}-\hspace{-0.7mm}\Gamma_1) \lambda_2 \lambda_S + (1\hspace{-1mm}-\hspace{-0.7mm}\Gamma_2) \lambda_1 \lambda_S + (\Gamma_1\hspace{-1mm}+\hspace{-0.7mm}\Gamma_2) \lambda_1 \lambda_2)   +   (1\hspace{-1mm}-\hspace{-0.7mm}\Gamma_1)(1\hspace{-1mm}-\hspace{-0.7mm}\Gamma_2) \lambda_S + \Gamma_1(1\hspace{-1mm}-\hspace{-0.7mm}\Gamma_2) \lambda_1   +   (1\hspace{-1mm}-\hspace{-0.7mm}\Gamma_1)\Gamma_2 \lambda_2} \label{eq:second_best_I_B1} \\
			I_2^{sb} & = & \frac{E[\theta_2-\theta_S] (1-\Gamma_2) b \lambda_1 \lambda_S   -   E[\theta_2-\theta_S](1-\Gamma_1)(1-\Gamma_2) \lambda_S   -   E[\theta_2-\theta_1] \Gamma_1 (1-\Gamma_2) \lambda_1}{b^2 \lambda_1 \lambda_2 \lambda_S - b ((1\hspace{-1mm}-\hspace{-0.7mm}\Gamma_1) \lambda_2 \lambda_S + (1\hspace{-1mm}-\hspace{-0.7mm}\Gamma_2) \lambda_1 \lambda_S + (\Gamma_1\hspace{-1mm}+\hspace{-0.7mm}\Gamma_2) \lambda_1 \lambda_2)   +   (1\hspace{-1mm}-\hspace{-0.7mm}\Gamma_1)(1\hspace{-1mm}-\hspace{-0.7mm}\Gamma_2) \lambda_S + \Gamma_1(1\hspace{-1mm}-\hspace{-0.7mm}\Gamma_2) \lambda_1   +   (1\hspace{-1mm}-\hspace{-0.7mm}\Gamma_1)\Gamma_2 \lambda_2} \label{eq:second_best_I_B2} \\
			q_1^{sb} & = & \frac{\theta_1\hspace{-1mm}-\hspace{-0.7mm}\theta_S}{b} \hspace{-1mm}+\hspace{-0.7mm} \frac{E[\theta_1\hspace{-1mm}-\hspace{-0.7mm}\theta_S] \big( b \lambda_2 ((1\hspace{-1mm}-\hspace{-0.7mm}\Gamma_1)\lambda_S \hspace{-1mm}+\hspace{-0.7mm} \Gamma_1 \lambda_1) \hspace{-1mm}-\hspace{-0.7mm} \Gamma_1 (1\hspace{-1mm}-\hspace{-0.7mm}\Gamma_2) \lambda_1 \hspace{-1mm}-\hspace{-0.7mm}(1\hspace{-1mm}-\hspace{-0.7mm}\Gamma_1)\Gamma_2 \lambda_2 \hspace{-1mm}-\hspace{-0.7mm}(1\hspace{-1mm}-\hspace{-0.7mm}\Gamma_1)(1\hspace{-1mm}-\hspace{-0.7mm}\Gamma_2) \lambda_S \big)   \hspace{-1mm}+\hspace{-0.7mm}   E[\theta_2\hspace{-1mm}-\hspace{-0.7mm}\theta_S] \Gamma_2 b \lambda_1 \lambda_2}{b \big( b^2 \lambda_1 \lambda_2 \lambda_S \hspace{-1mm}-\hspace{-0.7mm} b ((1\hspace{-1mm}-\hspace{-0.7mm}\Gamma_1) \lambda_2 \lambda_S \hspace{-1mm}+\hspace{-0.7mm} (1\hspace{-1mm}-\hspace{-0.7mm}\Gamma_2) \lambda_1 \lambda_S \hspace{-1mm}+\hspace{-0.7mm} (\Gamma_1\hspace{-1mm}+\hspace{-0.7mm}\Gamma_2) \lambda_1 \lambda_2)   \hspace{-1mm}+\hspace{-0.7mm}   (1\hspace{-1mm}-\hspace{-0.7mm}\Gamma_1)(1\hspace{-1mm}-\hspace{-0.7mm}\Gamma_2) \lambda_S \hspace{-1mm}+\hspace{-0.7mm} \Gamma_1(1\hspace{-1mm}-\hspace{-0.7mm}\Gamma_2) \lambda_1   \hspace{-1mm}+\hspace{-0.7mm}   (1\hspace{-1mm}-\hspace{-0.7mm}\Gamma_1)\Gamma_2 \lambda_2 \big)} \label{eq:second_best_q_1} \\
			q_2^{sb} & = & \frac{\theta_2\hspace{-1mm}-\hspace{-0.7mm}\theta_S}{b} \hspace{-1mm}+\hspace{-0.7mm} \frac{E[\theta_2\hspace{-1mm}-\hspace{-0.7mm}\theta_S] \big( b \lambda_1 ((1\hspace{-1mm}-\hspace{-0.7mm}\Gamma_2)\lambda_S \hspace{-1mm}+\hspace{-0.7mm} \Gamma_2 \lambda_2) \hspace{-1mm}-\hspace{-0.7mm} (1\hspace{-1mm}-\hspace{-0.7mm}\Gamma_1) \Gamma_2 \lambda_2 \hspace{-1mm}-\hspace{-0.7mm}(1\hspace{-1mm}-\hspace{-0.7mm}\Gamma_2)\Gamma_1 \lambda_1 \hspace{-1mm}-\hspace{-0.7mm}(1\hspace{-1mm}-\hspace{-0.7mm}\Gamma_1)(1\hspace{-1mm}-\hspace{-0.7mm}\Gamma_2) \lambda_S \big)   \hspace{-1mm}+\hspace{-0.7mm}   E[\theta_1\hspace{-1mm}-\hspace{-0.7mm}\theta_S] \Gamma_1 b \lambda_1 \lambda_2}{b \big( b^2 \lambda_1 \lambda_2 \lambda_S \hspace{-1mm}-\hspace{-0.7mm} b ((1\hspace{-1mm}-\hspace{-0.7mm}\Gamma_1) \lambda_2 \lambda_S \hspace{-1mm}+\hspace{-0.7mm} (1\hspace{-1mm}-\hspace{-0.7mm}\Gamma_2) \lambda_1 \lambda_S \hspace{-1mm}+\hspace{-0.7mm} (\Gamma_1\hspace{-1mm}+\hspace{-0.7mm}\Gamma_2) \lambda_1 \lambda_2)   \hspace{-1mm}+\hspace{-0.7mm}   (1\hspace{-1mm}-\hspace{-0.7mm}\Gamma_1)(1\hspace{-1mm}-\hspace{-0.7mm}\Gamma_2) \lambda_S \hspace{-1mm}+\hspace{-0.7mm} \Gamma_1(1\hspace{-1mm}-\hspace{-0.7mm}\Gamma_2) \lambda_1   \hspace{-1mm}+\hspace{-0.7mm}   (1\hspace{-1mm}-\hspace{-0.7mm}\Gamma_1)\Gamma_2 \lambda_2 \big)} \label{eq:second_best_q_2} 
		\end{eqnarray}
		\end{small}
		\begin{small}		
		\begin{eqnarray}
			\Pi_{HQ}^{sb} & = & E[\theta_1-\theta_S]^2 \big( b^2(\lambda_1 \lambda_2 \lambda_S^2 + \lambda_1 \lambda_2^2 \lambda_S + \lambda_1^2 \lambda_2 \lambda_S) - b(\lambda_1 \lambda_S^2 + 3 \lambda_1 \lambda_2 \lambda_S + \lambda_1^2 \lambda_S) + 2 \lambda_1 \lambda_S) \big)   +  \nonumber \\
			& & E[\theta_2-\theta_S]^2 \big( b^2(\lambda_1 \lambda_2 \lambda_S^2 + \lambda_1 \lambda_2^2 \lambda_S + \lambda_1^2 \lambda_2 \lambda_S) - b(\lambda_2 \lambda_S^2 + 3 \lambda_1 \lambda_2 \lambda_S + \lambda_2^2 \lambda_S) + 2 \lambda_2 \lambda_S) \big)   -  \nonumber \\ \label{eq:second_best_HQ}
			& & E[\theta_1-\theta_2]^2 \big( b(\lambda_1 \lambda_2^2 + \lambda_1^2 \lambda_2) - 2 \lambda_1 \lambda_2 \big) / \\
			& & 2\big( b^3 (\lambda_1 \lambda_2 \lambda_S^2 + \lambda_1 \lambda_2^2 \lambda_S + \lambda_1^2 \lambda_2 \lambda_S) -b^2(\lambda_1 + \lambda_2 + \lambda_S)(\lambda_1 \lambda_S + 2\lambda_1 \lambda_2 +\lambda_2 \lambda_S) +  \nonumber \\
			& & b((\lambda_1 + \lambda_2 + \lambda_S)^2 +4 \lambda_1 \lambda_2 +\lambda_1 \lambda_S + \lambda_2 \lambda_S) -2(\lambda_1 + \lambda_2 + \lambda_S) \big) \nonumber
		\end{eqnarray}
		\end{small}
		
		\end{landscape}
		}

\newpage		
		\hspace{-6.5mm} Note, if $E[\theta_1]=E[\theta_2]$, $\lambda_S=1$, and $\lambda_1=\lambda_2=0.5$, then each surplus sharing parameter $\Gamma_j$ is one half, which means that the contribution margin $M_j$ achieved is divided between the supplying division and the $j$th buying division in equal shares. 
		But independently of how the headquarters determines $\Gamma_j$, each division tends to underinvest, since each division has to bear the investment costs on its own. 
		As a consequence, the first-best quantities resulting from the negotiation process cannot be reached either, which implies that the resulting headquarters' profit $\Pi_{HQ}^{sb}$ is smaller than the headquarters' profit in the first-best case $\Pi_{HQ}^*$. 
		
		A headquarters that has all the information required to solve the $\Gamma$-choice problem and knows exactly, how its divisions will operate, would choose $\Gamma_1^{sb}$ and $\Gamma_2^{sb}$ according to Eq. \ref{eq:Gamma_1_sb} and Eq. \ref{eq:Gamma_2_sb}, respectively. 
		But how could a headquarters anticipate how its divisions will respond in the future.
		A headquarters with a lack of knowledge therefore needs a rule of thumb or alternative approaches that are easier to implement in order to determine the optimal share of the contribution margin achieved between divisions. 
		










\section{Simulation model with fuzzy Q-learning agents}
\label{sec:agent_based_simulation} 

	\subsection{Overview of the simulation model}
	\label{ssec:Relaxed_Assumptions}

		The negotiated transfer pricing model described in the preceding section expects that all three divisions make decisions by means of their anticipated information. 
		However, in decentralized business organizations, it is common that not all parties have access to information which are required to make optimal decisions. 
		Therefore the common knowledge assumption is widely relaxed in the simulation model (see Tab. \ref{tab:Overview} for the most important adaptations regarding the common knowledge assumption and, correspondingly, how the decision variables are determined). 
		Since the headquarters applies incentive compatibilities for ex post efficient quantities (cf. Eq. \ref{eq:Profit_S} - \ref{eq:Profit_2}), all three divisions have an incentive to truthfully share their private information with each other during the negotiation process. 
		In the agentized version of the negotiated transfer pricing model, the headquarters' profit function $\Pi_{HQ}$ is known to each division, but the division's profit function $\Pi_j$ as well as the expected value of the state variable $E[\theta_j]$, for $j \in \{ S,1,2 \}$, remain private for each division. 
		
		\afterpage{
		\begin{landscape}
		\begin{table}[h!]
		\linespread{1.1}\selectfont
		\centering
		\caption{Comparison between negotiated transfer pricing with fully rational agents and the agent-based variant with cognitively bounded agents.}
		\label{tab:Overview}
		\begin{tabularx}{186mm}{|c|l|l|l|}
			\hline
			\multirow{2}{*}{Parameter} & \multirow{2}{*}{Description} & Negotiated transfer pricing  & Agent-based variant of negotiated transfer pricing \\
			& & with fully rational agents (see Sec. \ref{sec:The_Model}) & with cognitively bounded agents (see Sec. \ref{sec:agent_based_simulation}) \\
			\hline
			\rule{0pt}{9pt} $\Pi_{HQ}$ & headquarters' profit & common knowledge & common knowledge \\
			\rule{0pt}{12pt} $\Pi_j$ & division's profit & common knowledge & private information for entire duration \\
			\rule{0pt}{12pt} $E[\theta_j]$ & expected value of state variable & common knowledge & private information for entire duration \\
			\rule{0pt}{12pt} $C_j$ & division's costs & common knowledge & private information until negotiation \\
			\rule{0pt}{12pt} $R_j$ & division's net revenue & common knowledge & private information until negotiation \\
			\rule{0pt}{12pt} $\lambda_j$ & division's marginal cost & common knowledge & private information until negotiation \\
			\rule{0pt}{12pt} $w_j$ & division's investment costs & common knowledge & private information until negotiation \\
			\rule{0pt}{12pt} $b$ & slope of the inv. demand func. & common knowledge & private information until negotiation \\
			\hline
			\rule{0pt}{9pt} $\Gamma_j$ & surplus sharing parameter & is set optimally & is a scenario-based exogenous parameter, which \\
			& & & varies in small steps \\
			\rule{0pt}{12pt} $I_j$ & amount of specific investment & is set optimally given $\Gamma_j$ & is chosen by an exploration policy, which mainly \\
			& & & depends on the learned Q-function \\
			\rule{0pt}{12pt} $\theta_j$ & state variable & private information until negotiation & private information until negotiation \\
			\rule{0pt}{12pt} $q_j$ & quantity & is set optimally given $I_j$ and $\theta_j$ & is set optimally given $I_j$ and $\theta_j$ \\
			\hline
		\end{tabularx}
		\linespread{1}\selectfont
		\end{table}
		\end{landscape}
		}
		
		In the agent-based simulation, the choice of different $\Gamma_j$ values, $j \in \{ 1,2\}$, is examined. 
		$\Gamma_j$ is a scenario-based exogenous parameter, which is varied in small steps (see Tab. \ref{tab:Parameter_settings_and_scenario_overview}), in order to investigate whether $\Gamma_j^{sb}$ resulting from a headquarters which has access to all required information leads to higher profit-effectiveness than other $\Gamma_j$ constellations taken from a headquarters which cannot solve the three-stage decision problem analytically due to a lack of knowledge and uses, e.g, a fifty-fifty surplus sharing rule. 
		In addition, the study assumes that each division has no beliefs about the rationality of other divisions nor how other divisions face their maximization problem. 
		Therefore, the divisions cannot apply the concept of subgame perfect equilibrium as in Sec. \ref{sec:The_Model} to solve the multi-stage decision-making process and, in particular, determine the optimal values for $I_j$ according to Eq. \ref{eq:second_best_I_S} - \ref{eq:second_best_I_B2}.  
		Thus, each division has first to learn which level of investment leads to which consequence and, in doing so, the study assumes that the divisions behave like fuzzy Q-learning agents. 
		Given that the divisions no longer instantaneously see the consequences of their actions, an exploration policy for the amount of specific investment $I_j$ has to be chosen, which is discussed in Sec. \ref{ss:Exploration_Polic} in more detail. 
		
		 
		Since the simulation study assumes that all divisions have no prior knowledge about the consequences of their decisions, the sequence of events within the negotiated transfer pricing model is run through several times. 
		This additional time dimension is characterized by a time subscript $t \in \mathbb{N}$, which describes the ``inner loop'' of the simulation.
		A flow diagram of the agent-based simulation is given in Fig. \ref{fig:Procedure} in Appendix A. 

\subsection{Fuzzy Q-learning}
\label{sec:The_learning_method}

			In this paper, fuzzy Q-learning proposed by \cite{glorennec1994fuzzy} is applied to describe the before-mentioned divisions, hereinafter referred to as agents.\footnote{
			Readers unfamiliar with reinforcement learning methods should refer to \cite{sutton2018reinforcement}. 
			For agent-based simulations with fuzzy Q-learning agents acting in an economic context see, e.g., \cite{kofinas2018fuzzy, rahimiyan2006modeling}.} 
			Note that fuzzy logic is based on the premise that the key elements in human thinking are not numbers, but labels of fuzzy sets \citep{zadeh1973outline}. 
			Since human thinking is tolerant of imprecision, fuzzy logic is suitable for representing human decision-making in a natural way \citep[e.g.,][]{zadeh1973outline, zimmermann2011fuzzy}. 
			By doing so, the fuzzy conditional statements are expressions of the form, IF $A$ THEN $B$, where $A$ and $B$ have fuzzy meaning. 
			The fuzzy rule-based system in this simulation study has the following form, which can be regarded as a zero order Takagi-Sugeno fuzzy system. 
			For each fuzzy rule $i$:
			\begin{align}
				\text{IF } s_{S,t} \text{ IS } L_S^i & \text{ AND } s_{1,t}\text{ IS } L_1^i \text{ AND } s_{2,t}\text{ IS } L_2^i \text{ THEN } \label{eq:fuzzy_rule_based_system} \\ 
				 & Q_{j,t}(\pmb{s}_{t}, \pmb{a}_{j,t}) = \sum\limits_{i=1}^N \alpha_i(\pmb{s}_t) \; q_{j,t}[i,a_{j,t,i}] \label{eq:Q_values} \\ 
				& A_{j,t}(\pmb{s}_{t}, \pmb{a}_{j,t}) = \sum\limits_{i=1}^N \alpha_i(\pmb{s}_t) \; a_{j,t}[i,a_{j,t,i}] \label{eq:A_values}
			\end{align}
			where the parameter $i \in \{ 1,...,N \}$ describes the index of fuzzy rules, $N \in \mathbb{N}$ is the number of fuzzy rules, $\pmb{s}_t = (s_{S,t},s_{1,t},s_{2,t}) \in \mathcal{S} \subset \mathbb{R}^3$ represents the state vector, $\pmb{a}_{j,t}=(a_{j,t,1},...,a_{j,t,N}) \in \{ 1,...,K \}^N$ denotes the index vector of stored actions, $a_{j,t}[i, a_{j,t,i}] \in \mathcal{A} \subset \mathbb{R}$ are the stored actions, $K \in \mathbb{N}$ is the number of stored actions in each fuzzy rule, $A_{j,t}(\pmb{s}_t, \pmb{a}_{j,t}) \in \mathbb{R}$ refers to the inferred action, and $q_{j,t}[i,a_{j,t,i}] \in \mathbb{R}$ denotes the stored q-values, whereby $a_{j,t}[., \hspace{-0.6mm} .]$ and $q_{j,t}[., \hspace{-0.6mm} .]$ are stored in look-up tables (indicated by square brackets).
			The number of fuzzy rules $N$ and the number of stored actions $K$ rely on the fuzzy partition of the state space $\mathcal{S}$ and on the discretization of the action space $\mathcal{A}$, respectively. 
			For the sake of simplicity, it is assumed that each fuzzy rule has exactly $K$ possible actions. 
			
			Note, in fuzzy Q-learning, $\mathcal{S}$ and $\mathcal{A}$ are continuous spaces and the inferred action $A_{j,t}(\pmb{s}_t, \pmb{a}_{j,t})$ of agent $j$ corresponds to the investment decision $I_{j,t}$, for $j \in \{ S,1,2 \}$. 
			Further, the fuzzy sets $L_j^i$ are characterized by linguistic labels and the function $\alpha_i(\pmb{s}_t)$ denotes the truth value of rule $i$ given the state vector $\pmb{s}_t$. 
			Q-values $Q_{j,t}(\pmb{s}_t, \pmb{a}_{j,t})$ and actions $A_{j,t}(\pmb{s}_t, \pmb{a}_{j,t})$ are inferred from Eq. \ref{eq:Q_values} and Eq. \ref{eq:A_values}, respectively. 
	
			The fuzzy rule-based system in this paper assumes that the ``weights" $\alpha_i(\pmb{s}_t)$ are generated by the T-norm product 
			\begin{equation}
				\alpha_i(\pmb{s}_t) = \mu_{L_S^i}(s_{S,t}) \cdot \mu_{L_1^i}(s_{1,t}) \cdot \mu_{L_2^i}(s_{2,t}) \;, \label{eq:truth_value} 
			\end{equation}
			where $\mu_{L_j^i}(s_{j,t}) \in [0,1]$ denotes the membership function (or membership grade) of rule $i$ in state $s_{j,t}$, $j \in \{ S,1,2 \}$. 
			The T-norm (or triangular norm, see, e.g., \cite{zimmermann2011fuzzy}) is a type of binary operation that is often used in fuzzy logic to model the AND operator in Eq. \ref{eq:fuzzy_rule_based_system}. 
			The membership functions considered here are defined on the interval $[0,1]$, which implies that if an object has a membership grade of one (zero) in a fuzzy set, then the object is absolutely (not) in that fuzzy set. 
			In addition, the membership functions are set in such a way that the strong fuzzy partition is fulfilled, i.e., $\sum_{i=1}^N \alpha_i(\pmb{s}_t) =1$ for each $\pmb{s}_t \in \mathcal{S}$. %
			
			Finally, the stored q-values are updated by 
			\begin{equation}
				q_{j,t+1}[i,a_{j,t,i}] = q_{j,t}[i,a_{j,t,i}] + \alpha_i(\pmb{s}_t) \; \Delta Q_{j,t}(\pmb{s}_t, \pmb{a}_{j,t}) \;,	\label{eq:Fuzzy_Q_Update_Rule} 	
			\end{equation}
			where $\Delta Q_{j,t}(\pmb{s}_t, \pmb{a}_{j,t})$ is the temporal difference error which is given by 
			\begin{equation}
				\alpha \Big( r_{j,t}(\pmb{s}_t,A_{j,t}(\pmb{s}_t, \pmb{a}_{j,t})) + \gamma \sum\limits_{i=1}^N \alpha_i(\pmb{s}_{t+1}) \hspace{-1mm} \max\limits_{k \in \{ 1,...,K\}} \hspace{-1mm} q_{j,t}[i,k] - Q_{j,t}(\pmb{s}_t, \pmb{a}_{j,t}) \Big) \; . \label{eq:Fuzzy_Q_Error} 	
			\end{equation}
			In Eq. \ref{eq:Fuzzy_Q_Error}, $r_{j,t} \in \mathbb{R}$ describes the reward of agent $j$ which is the division's profit $\Pi_j$ at time $t \in \mathbb{N}$, $j \in \{ S,1,2 \}$. 
			$\alpha \in (0,1]$ and $\gamma \in [0,1)$ denote the learning rate and the discount factor, respectively. 
			For the sake of simplicity, it is assumed that all three agents have the same learning rate as well as the same discount factor. 

	\subsection{Exploration policy} 
	\label{ss:Exploration_Polic}
	
		When using a reinforcement learning method, a trade-off is required between choosing the currently optimal action and choosing a varied action with the prospect of a higher reward in the future \citep{sutton2018reinforcement}. 
		Thus, the fuzzy Q-learning agents, which are situated in a non-stationary environment, need an exploration policy which ensures that all actions are performed frequently enough so that the fuzzy Q-function converges to an optimum. 
		In this simulation study, three commonly used exploration policies are applied to verify whether the results are robust to changes in the exploration policy. 





		\subsubsection{Boltzmann exploration policy}
		\label{sssec:Boltzmann_Exploration_Policy}
		
			The Boltzmann, Gibbs, or softmax exploration policy \citep[][]{cesa2017boltzmann} is a classic strategy for sequential decision-making under uncertainty. 
			The probability of choosing an action is proportional to an exponential function of the empirical mean of the reward of that action.
			The Boltzmann exploration policy is given by the probability mass function 
			\begin{equation}
				P_{j,t,i,k}(a_t[i,k]) = \frac{exp\big( {q_{j,t}[i,k]/\beta_j(t)} \big)}{\sum_{l=1}^{K} exp\big( {q_{j,t}[i,l]/\beta_j(t)} \big)} \;, \text{ for } k \in \{ 1,...,K\} \;, \label{eq:Boltzmann_Probability} 
			\end{equation}
			where $\beta_j(t) \in \mathbb{R}^+$, $j \in \{ S,1,2\}$, controls the degree of randomness for exploration \citep{powell2012optimal}. 
			Note that, for each agent $j$, for each time step $t$, and for each fuzzy rule $i$, it holds $\sum_{k=1}^{K} P_{j,t,i,k} = 1$ and, moreover, that actions with higher q-values are more likely to be selected than actions with lower q-values. 
			After all probabilities have been calculated according to Eq. \ref{eq:Boltzmann_Probability}, the index of each stored action is drawn from
			\begin{equation}
				a_{j,t,i} \thicksim \bigg( \genfrac{}{}{0pt}{0}{1}{P_{j,t,i,1}},...,\genfrac{}{}{0pt}{0}{K}{P_{j,t,i,K}} \bigg) \;. 
			\end{equation}

		\subsubsection{$\epsilon$-greedy exploration policy}
		\label{sssec:Epsilon_Greedy_Exploration_Policy}
		
			A very easy and commonly used exploration policy in reinforcement learning is the $\epsilon$-greedy exploration policy.
			It chooses the action with the highest q-value in the current state with probability $1- \epsilon$ and a random action otherwise \citep{tijsma2016comparing}. 
			The $\epsilon$-greedy exploration policy is given by
			\begin{equation}
				a_{j,t,i}= \begin{cases}
				\vspace{3mm} \underset{k \in \{ 1,...,K \}}{\mathrm{arg\,max}} q_{j,t}[i,k] & \text{with probability $1-\epsilon(t)$} \\
				\thicksim Unif \Big( \{ 1 ,...,K \} \Big) & \text{with probability $\epsilon(t)$}
				\end{cases}
			\end{equation}	
			where $\epsilon(t) \in [0,1]$ determines the degree of randomness for exploration which decreases over time. 
			Note that the $\epsilon$-greedy exploration policy can be seen as a benchmark for other more sophisticated exploration policies \citep{tijsma2016comparing}. 

		\subsubsection{Upper confidence bound exploration policy}
		\label{sssec:Upper_Confidence_Bound_Exploration_Policy}

			The third and last exploration policy is the so-called upper confidence bound exploration policy (hereafter abbreviated to UCB policy). 
			The idea of the UCB policy is that the square-root expression is a measure of the uncertainty in the estimate of the action in the current state \citep{sutton2018reinforcement}. 
			After choosing an action, the uncertainty is reduced by increasing the number of actions by one.
			For each agent $j \in \{ S,1,2 \}$, the UCB policy is given by 
			\begin{equation}
				a_{j,t,i}= \begin{cases}
				\vspace{3mm} \underset{k \in \{ 1,...,K \}}{\mathrm{arg\,max}} q_{j,t}[i,k] + c_1 \sqrt{\frac{ln(t)}{N_{j,t}[i,k]}} & \text{if $\forall k \; N_{j,t}[i,k] > 0$} \\ 
				\thicksim Unif \Big( \{ k \in \{ 1,...,K \} \; | \; N_{j,t}[i,k]=0 \} \Big) & \text{else} 
				\end{cases}
			\end{equation}
			where $c_1 \in \mathbb{R}$ controls the degree of randomness for exploration and $N_{j,t}[i,k] \in \mathbb{N}$ is the number of times that the index of the stored action $k$ in fuzzy rule $i$ has been chosen prior to time $t$ \citep[][]{sutton2018reinforcement}.  

\afterpage{
		\begin{landscape}
		\begin{table}[h!]
		\linespread{1.0}\selectfont
		\centering
		\caption{Parameter settings and scenario overview}
		\label{tab:Parameter_settings_and_scenario_overview}
		\begin{tabularx}{179mm}{|l|cccc|}
			\hline
			\textbf{Fixed exogenous parameters} & \multicolumn{4}{c|}{\textbf{Values}} \\
			\hline
			Number of simulation runs & \multicolumn{4}{l|}{$10,\hspace{-0.6mm}000$} \\
			Number of times steps per simulation run & \multicolumn{4}{l|}{$T=2,\hspace{-0.6mm}100$} \\
			Time steps to learn the Q-function & \multicolumn{4}{l|}{$T_L=2,\hspace{-0.6mm}000$} \\
			Time steps to evaluate the outcome & \multicolumn{4}{l|}{$T_E=100$} \\
			Slope of the inverse demand function & \multicolumn{4}{l|}{$b=12$} \\
			Expected values of the state variables & \multicolumn{4}{l|}{$E[\theta_S]=60$,  $E[\theta_1]=E[\theta_2]=E[\theta_B]=100$} \\
			Action space and state space & \multicolumn{4}{l|}{$\mathcal{A}=\{ 0, 5,..., 50 \}$ and $\mathcal{S}= \{ 0, 12.5,..., 50 \}$} \\
			Number of fuzzy rules & \multicolumn{4}{l|}{$N=125$} \\
			Number of stored actions in each fuzzy rule & \multicolumn{4}{l|}{$K=11$} \\
			Learning rate & \multicolumn{4}{l|}{$\alpha=0.5$} \\
			Boltzmann exploration policy & \multicolumn{4}{l|}{$\beta_{S,1}=49975$, $\beta_{1,1}=\beta_{2,1} = 24987.5$, $\beta_{S,2}=\beta_{1,2}=\beta_{2,2}=498.75$}  \\
			$\epsilon$-greedy and UCB exploration policy & \multicolumn{4}{l|}{$\epsilon_1=1+1/1999$, $\epsilon_2=-1/1999$, and $c_S=60$, $c_1=c_2=30$} \\
			\hline
			\hline
			\multirow{2}{*}{\textbf{Scenario-based exogenous parameters}} & \multicolumn{4}{c|}{\textbf{Symmetric scenarios}} \\
			 & \textbf{I} & \textbf{II} & \textbf{III} & \textbf{IV} \\
			\hline
			$\lambda_S$ marginal cost of the supplying division & 1 & 1 & 1 & 1 \\
			$\lambda_1$ marginal cost of the 1st buying division & \multirow{2}{*}{0.5} & \multirow{2}{*}{0.222} & \multirow{2}{*}{(0.5, 0.222)} & \multirow{2}{*}{(0.222, 0.262, 0.307, 0.361, 0.424, 0.5)} \\
			$\lambda_2$ marginal cost of the 2nd buying division & & & & \\
			$\Gamma_1$ surplus sharing parameter between ``S and 1'' & \multirow{2}{*}{0.5} & \multirow{2}{*}{0.25} & \multirow{2}{*}{(0.25, 0.30,..., 0.55)} & \multirow{2}{*}{(0.25, 0.30,..., 0.55)} \\
			$\Gamma_2$ surplus sharing parameter between ``S and 2'' & & & & \\
			$\gamma$ discount factor & (0, 0.9) & (0, 0.9) & (0, 0.1,..., 0.9) & (0, 0.1,..., 0.9) \\
			$\sigma$ standard deviation of the state variables & 0 & 0 & 0 & (0, 5, 10) \\
			Exploration policy & BM & BM & BM & (Greedy, UCB, BM) \\
			\hline
			\hline
			\multirow{2}{*}{\textbf{Scenario-based exogenous parameters}} & \multicolumn{4}{c|}{\textbf{Non-symmetric scenarios}} \\
			 & \multicolumn{2}{c}{\textbf{V}} & \multicolumn{2}{c|}{\textbf{VI}} \\
			\hline
			$\lambda_S$ marginal cost of the supplying division & \multicolumn{2}{c}{1} & \multicolumn{2}{c|}{1} \\
			$\lambda_1$ marginal cost of the 1st buying division & \multicolumn{2}{c}{(0.534, 0.621)} & \multicolumn{2}{c|}{(0.5, 0.534, 0.564, 0.590, 0.609, 0.621)} \\
			$\lambda_2$ marginal cost of the 2nd buying division & \multicolumn{2}{c}{(0.463, 0.301)} & \multicolumn{2}{c|}{(0.5, 0.463, 0.425, 0.385, 0.343, 0.301)} \\
			$\Gamma_1$ surplus sharing parameter between ``S and 1'' & \multicolumn{2}{c}{(0.5, 0.55,..., 0.75)} & \multicolumn{2}{c|}{(0.5, 0.550, 0.600, 0.650, 0.700, 0.750)} \\
			$\Gamma_2$ surplus sharing parameter between ``S and 2'' & \multicolumn{2}{c}{(0.5, 0.45,..., 0.25)} & \multicolumn{2}{c|}{(0.5, 0.450, 0.400, 0.350, 0.300, 0.250)} \\
			$\gamma$ discount factor & \multicolumn{2}{c}{(0, 0.1,..., 0.9)} & \multicolumn{2}{c|}{(0, 0.1,..., 0.9)} \\
			$\sigma$ standard deviation of the state variables & \multicolumn{2}{c}{0} & \multicolumn{2}{c|}{(0, 5, 10)} \\
			Exploration policy & \multicolumn{2}{c}{BM} & \multicolumn{2}{c|}{(Greedy, UCB, BM)} \\
			\hline
		\end{tabularx}
		\linespread{1}\selectfont
		\end{table}
		\end{landscape}
	}

\section{Parameter settings}
\label{sec:Parameter_Settings_and_Simulation_Setup}

	This simulation study focuses on settings in which the learning behavior of the three agents is modeled by fuzzy Q-learning. 
	It is assumed that each division has an individual constant marginal cost parameter $\lambda_j$, $j \in \{ S,1,2\}$. 
	For reasons of simplicity, $\lambda_S$ is set to one in all examined scenarios, while $\lambda_1$ and $\lambda_2$ are varied in small steps (the corresponding $\lambda_j$ values are listed in Tab. \ref{tab:Parameter_settings_and_scenario_overview}). 
	Besides $\lambda_j$, also the surplus sharing parameter $\Gamma_j$, $j \in \{ 1,2\}$ is a scenario-based exogenous parameter which is set in small increments. 
	In addition, the discount factor $\gamma$ is varied between $0$ and $0.9$ in steps of $0.1$. 
	Note that a discount factor of zero indicates that the agent is very myopic, which implies that the agent cannot observe future rewards, while, if $\gamma$ goes to one, the agent becomes more non-myopic, which means that the agent will place more emphasis on future rewards. 
	To describe the turbulence of the market environment in which the agents operate, the standard deviations of the state variables $\sigma$ are set to $0$, $5$, and $10$, which can be interpreted as a deterministic market environment, a market environment with minor stochastic fluctuations, and an environment with considerable volatility on the markets. 
	Additionally, the state variables $\theta_S$ and $\theta_j$, $j \in \{ 1,2\}$, are captured in a normally distributed random variable with mean $60$ and $100$, respectively. 
	Finally, three different exploration policies are applied to check the robustness of the simulation results.
	In the following, the parameter settings in Tab. \ref{tab:Parameter_settings_and_scenario_overview} are discussed in more detail. 
	
	In Eq. \ref{eq:buying_division_net_revenue}, the slope of the inverse demand function $b$ is set to $12$, as this results in that, in the first scenario examined, the first- and second-best solutions have integer solutions (see Tab. \ref{tab:Firstbest_Secondbest_Solutions} in Appendix A). 
	The choice of $\mathcal{A}$ and $\mathcal{S}$ is set at least in such a way that the agents can always find the first- and second-best solution in all studied scenarios. 
	In this sense, the stored actions $a_{j,t}[i,a_{j,t,i}] \in \mathcal{A}$ vary between $0$ and $50$ in steps of $5$, while the states $\pmb{s}_t \in \mathcal{S}$ range from $0$ to $50$ in steps of $12.5$. 
	Consequently, there are $11$ stored actions for each fuzzy rule $i \in \{ 1,...,N \}$, $5$ states, and the number of fuzzy rules $N$ is $5^3=125$ (the fuzzy partition of the state space raised to the power of the number of agents). 
	
	\begin{figure}[b!]
		\centering
		\includegraphics[width=0.85\textwidth]{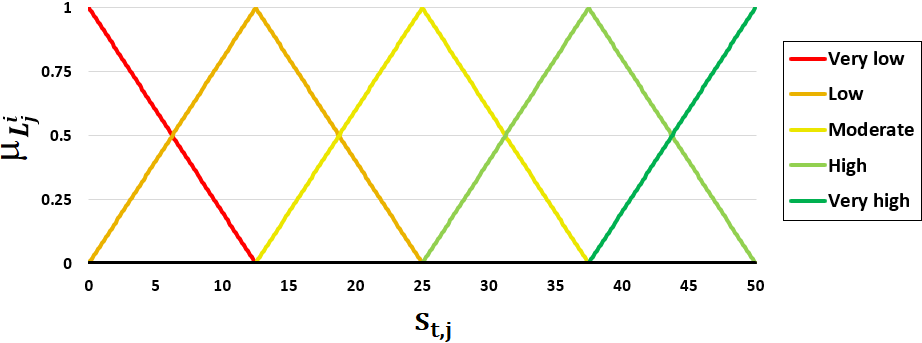}
		\caption{The fuzzy sets considered here consist of $5$ triangular membership functions in each state space dimension $j \in \{S,1,2\}$, which can be characterized by linguistic labels ranging from ``Very low'' to ``Very high''.}
		\label{fig:Triangular_Membership_Functions}
	\end{figure}	
	
	According to \cites{zimmermann2011fuzzy} suggestions, triangular membership functions are applied (cf. Fig. \ref{fig:Triangular_Membership_Functions}) because they are simple, have low computing power per iteration, and are easy to interpret. 
	For example, if the state of the supplying division $s_{t,S}$ is $10$, then the membership grade $\mu_{L_S^1}$ of fuzzy rule $1$ is $0.2$, whereas the membership grade $\mu_{L_S^2}$ of fuzzy rule $2$ is $0.8$, and all other membership grades $\mu_{L_S^3}$, $\mu_{L_S^4}$, and $\mu_{L_S^5}$ are zero, or in other words, the supplying division is in the ``very low investment range'' with a fraction of $20\%$ and in the ``low investment range'' with a share of $80\%$. 
	Incidentally, increasing the number of fuzzy rules means that the agents need more time to learn their environment.
	With only $5$ membership functions, the decision-making behavior can be exhibited quite well with the given parameter settings. 
	
	
	Furthermore, the learning rate $\alpha$ is set to $0.5$, because it can be assumed that, for long-term decisions, old stored information is just as important as new received information.  
	Regardless of the selection of the learning rate, the Q-function converges (sometimes faster and sometimes slower) to an optimum with the parameter settings giving in Tab. \ref{tab:Parameter_settings_and_scenario_overview}. 
	
	The number of time steps to learn the Q-function $T_L$ is set to $2,\hspace{-0.6mm}000$ as this is a sufficient number to guarantee that the Q-function converges in all investigated scenarios. 
	Note that this simulation study assumes that all agents have no prior knowledge of the consequences of their actions (all q-values are initialized to zero). 
	After determining the number of time steps required for learning the Q-function, further $100$ time steps are simulated to evaluate the agents' outputs, because, if $\sigma$ is not zero, the stochastic fluctuations of the markets affect the divisions' profits.
	Consequently, the number of time steps per simulation run $T$ is set to $2,\hspace{-0.6mm}100$.

	Since the choice of action is based on stochastic exploration policies, each scenario is carried out $10,\hspace{-0.6mm}000$ times and, due to the coefficient of variation (the ratio of the standard deviation to the mean), $10,\hspace{-0.6mm}000$ simulation runs are sufficient to express the precision and repeatability of the simulation study. 

	Finally, the used exploration policies are discussed. 
	Inspired by the work of \cite{tijsma2016comparing}, this simulation study focuses on the Boltzmann, $\epsilon$-greedy, and upper confidence bound exploration policy. 
	For the introduced Boltzmann exploration policy in Sec. \ref{sssec:Boltzmann_Exploration_Policy}, $\beta_j(t)$ should slowly decrease over time in order to reduce exploration \citep{dearden1998bayesian}. 
	In this sense, $\beta_j(t)$ is calculated by a simple rational function	
	\begin{equation}
		\beta_j(t) = \frac{\beta_{j,1}}{\beta_{j,2}+t} \; , \text{ for } j \in \{ S,1,2\}, \label{eq:experimentation_tendency}
	\end{equation}
	where $(\beta_{j,1},\beta_{j,2}) \in \mathbb{R}^2$ are set in such a way that the learning agents have a long time to explore for good policies, but also time to exploit them.\footnote{
	Technical note: The experimentation tendency $\beta_j(t)$ should not become too small during a simulation run, otherwise the expression $exp(q_{j,t}[i,a_{j,t,i}]/\beta_j(t))$ will be too large and this could endanger the numerical stability of the simulation.} 
	Note that rational functions, like Eq. \ref{eq:experimentation_tendency}, have the advantage over linear functions that the degree of randomness for exploration decreases faster at the beginning of a simulation run, but it never reaches zero. 
	According to pre-generated simulations, $\beta_S(t)$ should be about $100$ for the supplying division at the very beginning of a simulation run, while, at the end, a value of $20$ is sufficient for the convergence of the Q-function. 
	Conversely, for each buying division $j \in \{ 1,2\}$, $\beta_j(t)$ only needs to be half as large. 
	Solving Eq. \ref{eq:experimentation_tendency} with the two associated boundary conditions ($\beta_S(1)=100$ and $\beta_S(T_L)=20$ for the supplying division or rather $\beta_j(1)=50$ and $\beta_j(T_L)=10$ for each buying division $j \in \{ 1,2\}$) leads to the Boltzmann exploration parameters given in Tab. \ref{tab:Parameter_settings_and_scenario_overview}. 
	It should be mentioned that balancing the parameters of an exploration policy is one of the most challenging tasks in reinforcement learning \citep{tijsma2016comparing}. 
	
	For the $\epsilon$-greedy exploration policy (see Sec. \ref{sssec:Epsilon_Greedy_Exploration_Policy}), $\epsilon(t)$ is a simple decreasing linear function of the time
	\begin{equation}
		\epsilon(t) = \begin{cases}
		\vspace{3mm}  \epsilon_1 - \epsilon_2 \; t & \text{for $t \leq T_L$} \\
		0 & \text{for $t > T_L$}
		\end{cases}
	\end{equation}		
	with the two properties that, at the beginning of a simulation run, $\epsilon(1)$ is one, which indicates that the action-selection is total random (pure exploration) and, at the end, $\epsilon(T_L)$ is zero (pure exploitation). 
	The associated $\epsilon$-greedy exploration parameters $\epsilon_1$ and $\epsilon_2$ are reported in Tab. \ref{tab:Parameter_settings_and_scenario_overview}. 
	Finally, for the UCB exploration policy in Sec. \ref{sssec:Upper_Confidence_Bound_Exploration_Policy}, the pre-generated simulations reveal that a value of about $60$ is a good choice for $c_S$, while $c_1$ and $c_2$ only need to be half as large. 

\newpage
\section{Results and discussion}
\label{sec:Results_and_Discussion}

	The analysis of results is structured in two parts: 
	(1) results for symmetric marginal cost parameters $\lambda_1=\lambda_2$ (called symmetric scenarios) and (2) results for non-symmetric marginal cost parameters (called non-symmetric scenarios). 
	For a better overview, symmetric and non-symmetric scenarios are additionally divided into four and two sub-scenarios, respectively. 
	An overview of the examined scenarios is given in Tab. \ref{tab:Parameter_settings_and_scenario_overview}. 
	In the following, the supplying division, the 1st buying division, and the 2nd buying division are referred to as seller, first buyer, and second buyer, respectively. 
	In addition, this simulation study distinguishes between an all-knowing headquarters which sets the surplus sharing parameters according to Eq. \ref{eq:Gamma_1_sb} and Eq. \ref{eq:Gamma_2_sb}, and a headquarters which cannot solve the $\Gamma$-choice problem analytically and, therefore, chooses $\Gamma_j$, $j \in \{ 1,2 \}$, according to Tab. \ref{tab:Parameter_settings_and_scenario_overview}. 

	\subsection{Results for symmetric marginal cost parameters}

		In scenarios I-IV, $\lambda_S=1$ and $\lambda_1=\lambda_2$ which yields to $\Gamma_1=\Gamma_2$. 
		In scenarios I-III, the exploration policy is the Boltzmann exploration policy and the underlying market environment is deterministic. 
		Finally, scenario IV performs an extensive sensitivity analysis in order to verify whether the results for symmetric marginal cost parameters are robust to changes in volatility of the markets and exploration policy. 

	\subsubsection{Scenario I}
	
		In the first scenario, high investment costs are assumed on the downstream divisions. 
		Concretely, $\lambda_1$ and $\lambda_2$ are set to $0.5$ so that the simulation results can also be compared to \cites{mitsch2023impact} previous simulation study where only one buyer is involved. 
		If $\lambda_1=\lambda_2=0.5$, then an all-knowing headquarters would set $\Gamma_1$ and $\Gamma_2$ to $0.5$, according to Eq. \ref{eq:Gamma_1_sb} and Eq. \ref{eq:Gamma_2_sb}. 
		To get an understanding of the decision-making behavior of fuzzy Q-learning agents, the simulation model studies investment decisions with two extreme levels of no ($\gamma=0$) and high ($\gamma=0.9$) foresight of future rewards. 
		
		\begin{figure}[t!]
			\centering
			\includegraphics[width=0.942\textwidth]{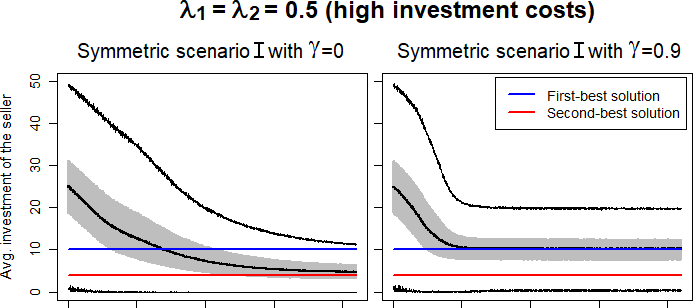} \\ 
			\includegraphics[width=0.5\textwidth]{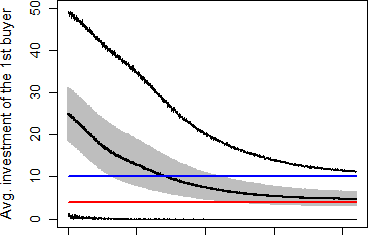} \includegraphics[width=0.434\textwidth]{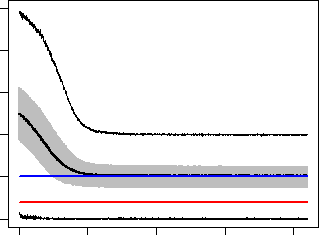} \\
			\includegraphics[width=0.5\textwidth]{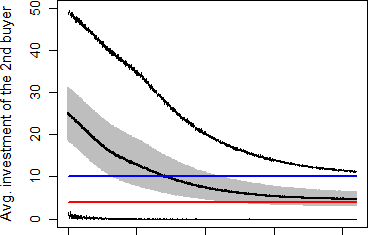} \includegraphics[width=0.434\textwidth]{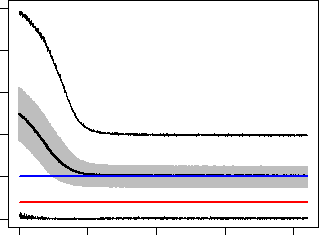} \\
			\includegraphics[width=0.5\textwidth]{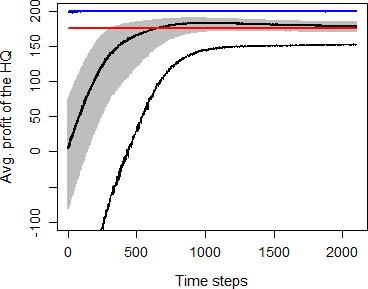} \includegraphics[width=0.434\textwidth]{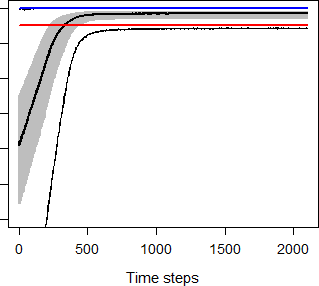}
			\caption{Modified boxplots of $I_S$, $I_1$, $I_2$, and $\Pi_{HQ}$ for myopic ($\gamma=0$) and non-myopic ($\gamma=0.9$) fuzzy Q-learning agents with $\lambda_S=1$, $\lambda_1=\lambda_2=0.5$, $\Gamma_1=\Gamma_2=0.5$, $\sigma=0$, and exploration policy is Boltzmann.}
			\label{fig:Plot_01}
		\end{figure}
		
		The simulation results of investment decisions and, what profits the headquarters can expect, are presented with modified boxplots (see Fig. \ref{fig:Plot_01}). 
		In addition, first- and second-best solutions are also displayed. 
		Be aware that the thick black line in the modified boxplots represents the median of $10,\hspace{-0.6mm}000$ observations per time step and, due to the fact that the mean value is not robust against outliers, the mean value of the headquarters' profit is below the median value in case of a negative skewness (roughly spoken, the mass of the distribution is concentrated on the ``right side''). 
		Additionally, note that, in the modified boxplots used here, the distance between the 25th and 75th percentile is visualized in gray, the outer black lines denote the whiskers, and, for an improved readability, outliers beyond the wiskers are not depicted. 
	
		For myopic fuzzy Q-learning agents (see left plots in Fig. \ref{fig:Plot_01}), the simulation results indicate that the investment decisions of all divisions converge towards the second-best solution. 
		In the case where all divisions seek for high future rewards (see right plots in Fig. \ref{fig:Plot_01}), the investment decisions converge towards the first-best solution. 
		Similarly, the headquarters' profit approaches the second-best solution and the first-best solution for myopic and non-myopic agents, respectively. 
		
		On looking more closely at the last $100$ investment decisions over $10,\hspace{-0.6mm}000$ simulation runs, the investment decisions of myopic (non-myopic) agents are normally distributed with mean $5.13$ ($10.02$) and standard deviation $0.44$ ($2.62$). 
		Moreover, the headquarters' profit for myopic (non-myopic) agents is nearly normally distributed with mean $177.9$ ($189.8$) and standard deviation $2.12$ ($6.37$), whereby the distribution of the headquarters' profit has a slightly negative skewness of $-0.21$ ($-0.98$) for myopic (non-myopic) agents. 
		
		The findings of scenario I show that myopic fuzzy Q-learning agents invest about as much as fully individual rational utility maximizers in the classic hold-up problem (cf. Tab. \ref{tab:Firstbest_Secondbest_Solutions} in Appendix A), whereas non-myopic agents invest optimally. 
		With a high level of foresight to strive for high future rewards, the investment level of each division increases on the one hand and the headquarters' profit on the other. 

	\subsubsection{Scenario II}
	
\begin{figure}[t!]
			\centering
			\includegraphics[width=0.942\textwidth]{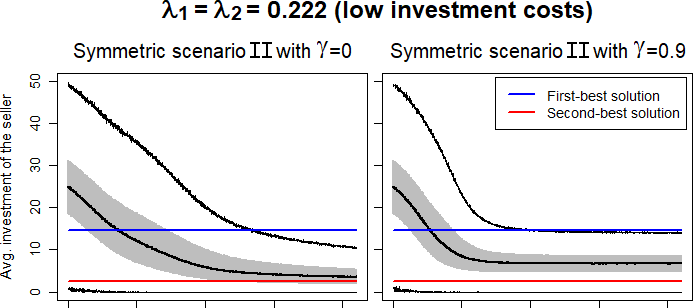} \\ 
			\includegraphics[width=0.5\textwidth]{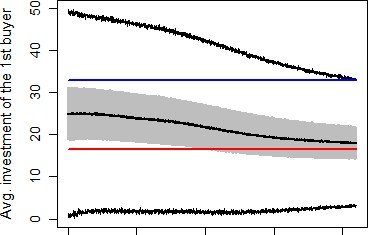} \includegraphics[width=0.434\textwidth]{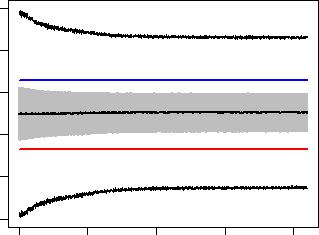} \\
			\includegraphics[width=0.5\textwidth]{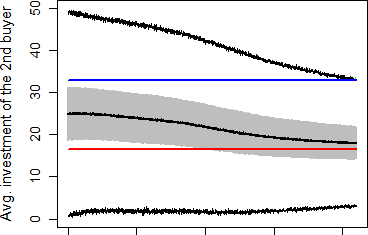} \includegraphics[width=0.434\textwidth]{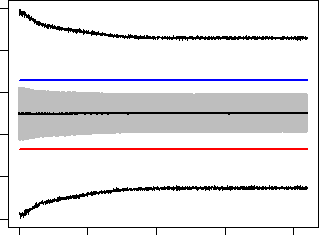} \\
			\includegraphics[width=0.5\textwidth]{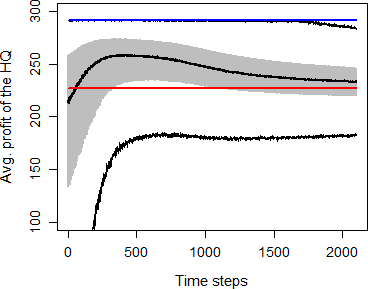} \includegraphics[width=0.434\textwidth]{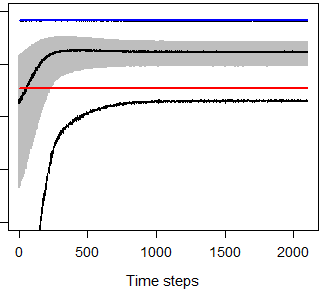}
			\caption{Modified boxplots of $I_S$, $I_1$, $I_2$, and $\Pi_{HQ}$ for myopic ($\gamma=0$) and non-myopic ($\gamma=0.9$) fuzzy Q-learning agents with $\lambda_S=1$, $\lambda_1=\lambda_2=0.222$, $\Gamma_1=\Gamma_2=0.25$, $\sigma=0$, and exploration policy is Boltzmann.}
			\label{fig:Plot_03}
		\end{figure}

		In the second scenario, the marginal cost parameters $\lambda_j$, $j \in  \{ 1,2\}$, are set to $0.222$. 
		In the case that both buying divisions have low investment costs, an all-knowing headquarters would choose $\Gamma_1 = \Gamma_2 = 0.25$. 
		The simulation results are depicted in the left plot for myopic agents and in the right plot for non-myopic agents in Fig. \ref{fig:Plot_03}. 
		As the output shows, myopic fuzzy Q-learning agents achieve second-best results, while non-myopic agents realize better results, but the first-best solution is not reached. 
		
		A closer analysis reveals that the investment decisions of myopic (non-myopic) sellers are nearly normally distributed with mean $4.27$ ($6.82$) and standard deviation $0.5$ ($1.78$) with a positive (negative) skewness of $0.48$ ($-0.25$). 
		On the downside, the investment decisions of myopic (non-myopic) buyers are normally distributed with mean $18.41$ ($25.24$) and standard deviation $1.53$ ($5.02$). 
		In the case of myopic (non-myopic) agents, the headquarters' profit is (almost) normally distributed with mean $233.7$ ($258.5$) and standard deviation $3.58$ ($11.5$) with a negative skewness of $-0.02$ ($-0.63$). 
		
		The findings of scenario II indicate that divisions' investment decisions of myopic fuzzy Q-learning agents are close to the second-best solutions (cf. Tab. \ref{tab:Firstbest_Secondbest_Solutions} in Appendix A), but compared to scenario I, non-myopic agents generate lower profits for themselves as well as for their headquarters. 
		A high degree of foresight favors divisions' investment decisions, however, on average, first-best investment decisions are not made. 

	\subsubsection{Scenario III}
	
\begin{figure}[t!]
		\centering
		\includegraphics[width=1\textwidth]{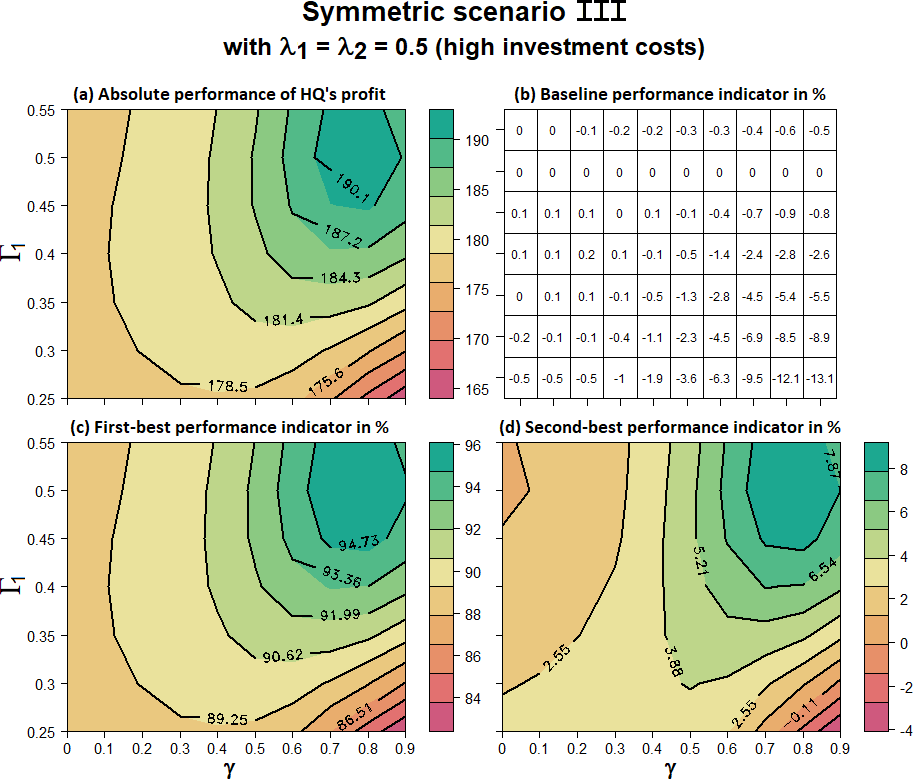}
		\caption{Results for symmetric marginal cost parameters with $\sigma=0$ and exploration policy is Boltzmann: (a) absolute performance of the headquarters' profit resulting from the simulation, (b) relative performance change and statistical hypothesis testing of $\Pi_{HQ}$ compared to the simulation output from the baseline scenario with $\Gamma_1^{sb}=0.5$, (c) relative performance of $\Pi_{HQ}$ compared to $\Pi_{HQ}^*$, and (d) relative performance change of $\Pi_{HQ}$ compared to $\Pi_{HQ}^{sb}$.} 
		\label{fig:Plot_05}
	\end{figure}

\begin{figure}[b!]
		\centering
		\includegraphics[width=1\textwidth]{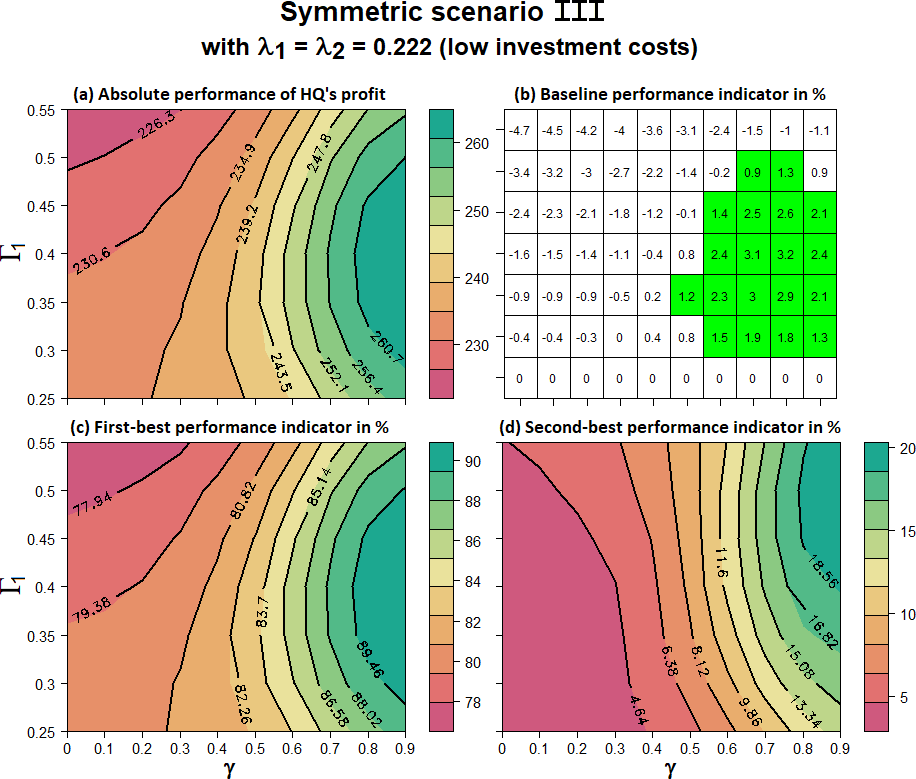}
		\caption{Results for symmetric marginal cost parameters with $\sigma=0$ and exploration policy is Boltzmann: (a) absolute performance of the headquarters' profit resulting from the simulation, (b) relative performance change and statistical hypothesis testing of $\Pi_{HQ}$ compared to the simulation output from the baseline scenario with $\Gamma_1^{sb} \hspace{-0.75mm} =0.25$, (c) relative performance of $\Pi_{HQ}$ compared to $\Pi_{HQ}^*$, and (d) relative performance change of $\Pi_{HQ}$ compared to $\Pi_{HQ}^{sb}$.}
		\label{fig:Plot_06}
	\end{figure}

		In order to get a general understanding of the effects of different values of surplus sharing parameters and discount factors, $\Gamma_j$ and $\gamma_j$, $j \in \{ 1,2\}$, are varied from $0.25$ to $0.55$ in $0.05$ increments and from $0$ to $0.9$ in $0.1$ increments, respectively.   	
		Note that, in the symmetric case, it is sufficient to vary $\Gamma_j, j \in \{ 1,2\}$, from $0.25$ to $0.55$ to adequately study the divisions' decision-making behavior. 
		The simulation results are summarized in Fig. \ref{fig:Plot_05} and Fig. \ref{fig:Plot_06} with four subplots for $\lambda_1=\lambda_2=0.5$ (high investment costs) and for $\lambda_1=\lambda_2=0.222$ (low investment costs), respectively. 
		In the following, the meaning of the subplots is explained. 
	
		The top left subplot displays contours representing the headquarters' profits resulting from the simulation. 
		The bottom left subplot depicts the ``first-best performance indicator'' for fuzzy Q-learning agents, which is defined by $\Pi_{HQ} / \Pi_{HQ}^*$, i.e., the headquarters' profit obtained from the simulation is normalized by the highest feasible profit that can be achieved. 
		The first-best performance indicator can serve as a relative indicator for the profit-effectiveness of the decision-making behavior of fuzzy Q-learning agents; the higher the value, the better the profitability for the headquarters. 
		The bottom right subplot shows the ``second-best performance indicator'', which reflects how much better fuzzy Q-learning agents perform than fully rational utility maximizers. 
		This second-best performance indicator is formalized by the relative change between $\Pi_{HQ}$ and $\Pi_{HQ}^{sb}$ dividing by $\Pi_{HQ}^{sb}$; the higher the relative change, the higher the performance of fuzzy Q-learning agents. 
		
		Lastly, and more importantly, the so-called ``baseline performance indicator'', which is displayed in the top right subplot. 
		The baseline performance indicator is based on the relative change between $\Pi_{HQ}^{\Gamma_j}$ and $\Pi_{HQ}^{baseline}$ dividing by $\Pi_{HQ}^{baseline}$, where $\Pi_{HQ}^{\Gamma_j}$ denotes the headquarters' profit resulting from scenario where the headquarters uses $\Gamma_j$, while $\Pi_{HQ}^{baseline}$ describes the headquarters' profit resulting from scenario where the headquarters applies $\Gamma_j^{sb}$. 
		The scenario in which the headquarters uses the optimal surplus sharing parameter $\Gamma_j^{sb}$ is called the ``baseline scenario''. 
		Note that in a baseline scenario, the headquarters knows all information from its divisions to determine the optimal level of bargaining power and, therefore, the headquarters can set the bargaining power $\Gamma_1$ and $\Gamma_2$ according to Eq. \ref{eq:Gamma_1_sb} and \ref{eq:Gamma_2_sb}, respectively. 
		So, this indicator provides information about how much better fuzzy Q-learning agents given $\Gamma_j$ perform than fuzzy Q-learning agents given $\Gamma_j^{sb}$ or, in other words, whether the profit of a headquarters that does not know the exact optimal surplus sharing parameters is higher than the profit of a headquarters that applies the optimal surplus sharing parameters. 

\newpage
		In order to find out whether the baseline performance indicator is significant (i.e., if there is a significant difference between $\Pi_{HQ}^{\Gamma_j}$ and $\Pi_{HQ}^{baseline}$), Welch's t-tests and Wilcoxon rank-sum tests are applied.\footnote{
		\cites{welch1947generalization} t-test and Wilcoxon rank-sum test \citep{mann1947test} are widely common statistical hypothesis tests.
		In the default case, the first one tests the equality of the means for two independent normally distributed samples with unequal and unknown variances, while the second one, a nonparametric test, tests the null hypothesis that the distributions of two independent samples differ by a location shift of zero.} 
		Since the p-values of hypothesis tests go quickly to zero when very large samples ($10,\hspace{-0.6mm}000$ observations and more) are evaluated, the null hypotheses become statistically significant because the standard errors become extremely small \citep{lin2013research}. 
		To mitigate this phenomenon, one-tailed Welch's t-tests and Wilcoxon rank-sum tests are used with a positive hypothesized mean difference (hereafter abbreviated to $d_H$). 
		The null hypotheses to be tested are given by $mean(\Pi_{HQ}^{\Gamma_j})-mean(\Pi_{HQ}^{baseline}) \geq d_H$.
		Be aware that the Welch's t-test is designed for normally distributed samples but, due to the central limit theorem, it can be assumed that this also holds for the headquarters' profits which are nearly normally distributed. 

		In order to find out, if the headquarters' profits differ significantly in terms of their means (see Welch's t-tests) or their distributions (see Wilcoxon rank-sum tests), the hypothesized mean difference $d_H$ is set to $\Pi_{HQ}^{baseline} / 100$, which can be interpreted as $1\%$ of the headquarters' profit achieved in the baseline scenario. 
		Therefore, whenever the t-test result is statistically significant, fuzzy Q-learning agents perform at least $1\%$ better in scenario with $\Gamma_j$ than fuzzy Q-learning agents in the baseline scenario with $\Gamma_j^{sb}$ given a standard significance level of $0.05$. 
		To better distinguish the test results graphically, the baseline performance indicator is colored as follows: If the p-value of the t-test (rank-sum test) is greater than $0.05$, the cell turns yellow (blue). 
		If both tests are statistically significant, the cell is green; otherwise the cell is white. 
		
		The simulation results on the performance of fuzzy Q-learning agents with $\lambda_1=\lambda_2=0.5$ and $\lambda_1=\lambda_2=0.222$ are presented in Fig. \ref{fig:Plot_05} and Fig. \ref{fig:Plot_06}, respectively. 
		According to subplot (a) in Fig. \ref{fig:Plot_05}, non-myopic fuzzy Q-learning agents tend to generate higher profits for their headquarters compared to myopic fuzzy Q-learning agents. 
		In cases where seller and buyer do not have equal bargaining power, i.e., $\Gamma_j \neq 0.5$, $j \in \{ 1,2 \}$, the headquarters' profit decreases. 
		In the extreme case with a non-symmetrical $\Gamma$-surplus sharing rule, $\Gamma_j=0.25$, $j \in \{ 1,2 \}$, non-myopic agents perform worse than myopic agents. 
		
		The efficiency of fuzzy Q-learning agents can be seen in the bottom contour plots (c) and (d). 
		According to subplot (c) in Fig. \ref{fig:Plot_05}, the first-best performance indicator shows that the profit-effectiveness of the headquarters is relatively high, especially for non-myopic agents. 
		The second-best performance indicator on the other side points out that fuzzy Q-learning agents slightly outperform fully rational utility maximizers in all investigated scenarios. 
		This results from the agents' learning phase in which learning begins without prior knowledge of the consequences of agents' actions (all q-values are initialized to zero). 
		The timelines in all modified boxplots show the same picture: Agents start with a moderate level of investment, which slowly decreases to its new equilibrium level (sometimes faster and sometimes slower). 
		When learning the Q-values, the agents will get higher rewards as they make higher investments. 
		But on the other hand, if too high investments are made, profits will decrease. 
		Also, if only one-sided investments are made, the agent's own profit is reduced while the other agents benefit from these investments. 
		Overall, fuzzy Q-learning agents tend to invest more than the benchmark of the second-best solution indicates. 
		Note that pre-generated simulations reveal that, regardless of the starting level of investments, the Q-functions converge to their final values which are slightly higher than the expected values resulting from fully rational utility maximizers. 
		
		Finally, to test whether the headquarters' profit is significantly different for fuzzy Q-learning agents with $\Gamma_j$ or $\Gamma_j^{sb}$, the baseline performance indicator is calculated. 
		Based on the baseline performance indicator (see subplot (b) in Fig. \ref{fig:Plot_05}), it may be inferred that, for symmetric surplus sharing parameters, $\Gamma_j=0.5$, $j \in \{ 1,2 \}$, a deviation from the recommended optimal solutions of subgame perfection equilibrium leads to worse results. 
		Therefore, the headquarters should follow the theory of subgame perfection equilibrium and should give all divisions the same bargaining power, or in other words, the headquarters should use a fifty-fifty surplus sharing rule. 
		
		The simulation results for $\lambda_1=\lambda_2=0.222$ (low investment costs) show a similar picture. 
		The absolute performance of the headquarters' profit rises with increasing degree of agents' foresight (cf. subplot (a) in Fig. \ref{fig:Plot_06}). 
		First- and second-best performance indicators indicate that fuzzy Q-learning agents generate relatively high profits, although, again, a high level of foresight favors agents' investment decisions. 
		According to subplot (b) in Fig. \ref{fig:Plot_06}, the results indicate that, for agents with a low discount factor $\gamma$ between $0$ and $0.4$, the headquarters should set $\Gamma_j$, $j \in \{ 1,2 \}$, to $\Gamma_j^{sb}$ for maximum profits. 
		However, for agents with a high foresight greater than $0.4$, the headquarters should empower the seller with greater bargaining power in order to achieve higher profits. 
		A closer look reveals that, in $19$ cases, Welch's t-test and Wilcoxon rank-sum test show a significant difference between $\Pi_{HQ}^{\Gamma_j}$ and $\Pi_{HQ}^{baseline}$, when the seller obtains a bargaining power which is greater than the recommended optimal surplus sharing rule resulting from solving the subgame perfect equilibrium. 
		
		In order to get a rule of thumb for determining the ideal surplus sharing rule, the weighted arithmetic mean of $\Gamma_j$ may be used. 
		For $j \in \{ 1,2 \}$, the weighted arithmetic mean of $\Gamma_j$ is given by the double series over all pairs of $\Gamma_j$ and $\gamma$, where the baseline performance indicator $BPI$ is significant: $\sum_{\Gamma_j=0.25}^{0.55} \sum_{\gamma=0}^{0.9} \Gamma_j \cdot BPI[\Gamma_j,\gamma] / \sum_{\Gamma_j=0.25}^{0.55} \sum_{\gamma=0}^{0.9} BPI[\Gamma_j,\gamma]$. 
 		According to subplot (b) in Fig. \ref{fig:Plot_06}, the weighted arithmetic mean of $\Gamma_j$ is $0.386$, which can be seen as a good choice for agents with a high degree of foresight between $0.5$ and $0.9$. 
		An $1:2$ surplus sharing rule for seller and buyer may therefore be a good choice for non-myopic agents when the headquarters cannot determine $\Gamma_j$ analytically. 

	\subsubsection{Scenario IV}
	
		To verify whether the simulation results are robust in terms of volatility on the markets and the implementation of other exploration policies, an extensive sensitivity analysis is performed (see figures in Appendix B). 
		To evaluate the robustness of stochastic fluctuations on the markets, the standard deviations of the state variables $\sigma$ vary from $0$ to $10$ in $5$ increments. 
		In addition to the Boltzmann exploration policy, the $\epsilon$-greedy and
the upper confidence bound exploration policy, which are introduced in Sec.
\ref{ss:Exploration_Polic}, are applied. 

\begin{figure}[h!]
		\centering
		\includegraphics[width=0.96\textwidth]{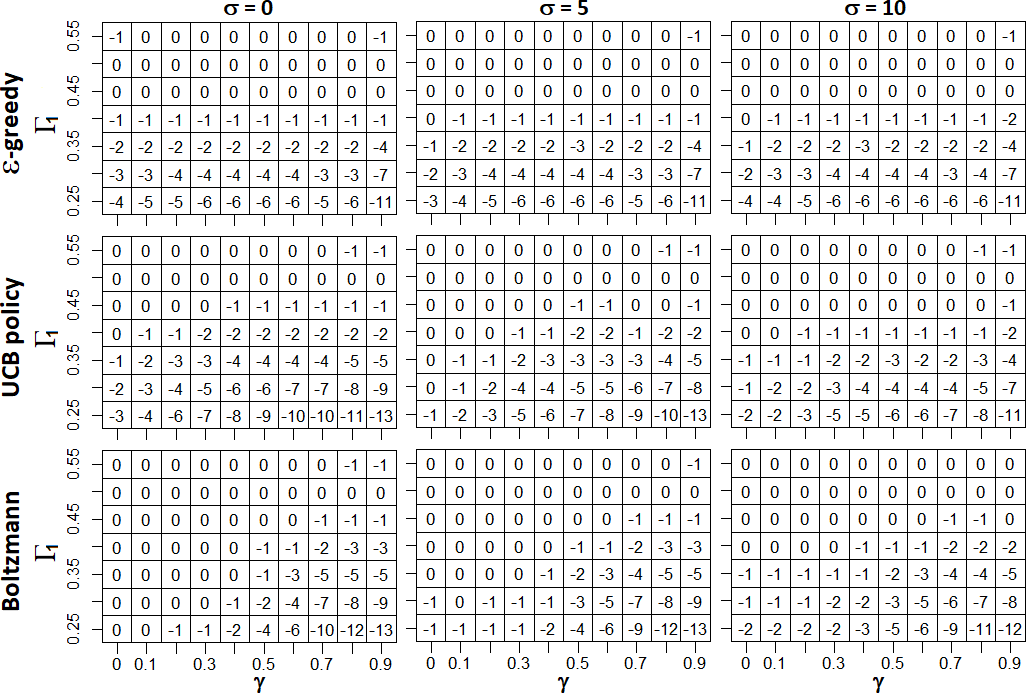}
		\caption{Results of the baseline performance indicator for $\Gamma_j^{sb}=0.5$ of different exploration policies and standard deviations of state variables given $\lambda_S=1$ and $\lambda_1=\lambda_2=0.5$.}
		\label{fig:Plot_08}
	\end{figure}

	In the case of symmetric marginal cost parameters, the learning behavior of fuzzy Q-learning agents seems to be robust against volatilities on the markets. 
	Also, all three exploration policies show a similar picture, whereby fuzzy Q-learning agents using the $\epsilon$-greedy exploration policy performing worst while agents applying Boltzmann performing best. 
	In scenarios where $\lambda_1=\lambda_2=0.5$ (see Fig. \ref{fig:Plot_08}), a symmetric $\Gamma$-surplus sharing rule leads to high profits, a deviation from it results in lower profits and should therefore not be pursued.  
	More importantly, the simulation results for $\lambda_1=\lambda_2<0.361$ (low investment costs) suggest that the seller should have a greater bargaining power in order to increase the headquarters' profit, especially when agents have a high foresight of future rewards. 
	In such cases, the headquarters should allocate the distribution ratio of the bargaining power at a ratio of $1:2$ for seller and buyer. 

	\subsection{Results for non-symmetric marginal cost parameters}
	
		Again, $\lambda_S$ is set to one, the exploration policy is the Boltzmann exploration policy, and the underlying market environment is deterministic. 
		In scenario V, the marginal cost parameter $\lambda_1$ is systematically increased from $0.534$ to $0.621$ in small increments, while $\lambda_2$ is decreased from $0.463$ to $0.301$ (cf. Tab. \ref{tab:Parameter_settings_and_scenario_overview}). 
		Also the surplus sharing parameters $\Gamma_1$ and $\Gamma_2$ differ accordingly. 
		Conclusively, scenario VI conducts an extensive sensitivity analysis in order to verify whether the results for non-symmetric marginal cost parameters are robust to changes in volatility of the markets and exploration policy. 
	
		\vspace{-2mm}
		\begin{figure}[b!]
			\centering
			\includegraphics[width=1\textwidth]{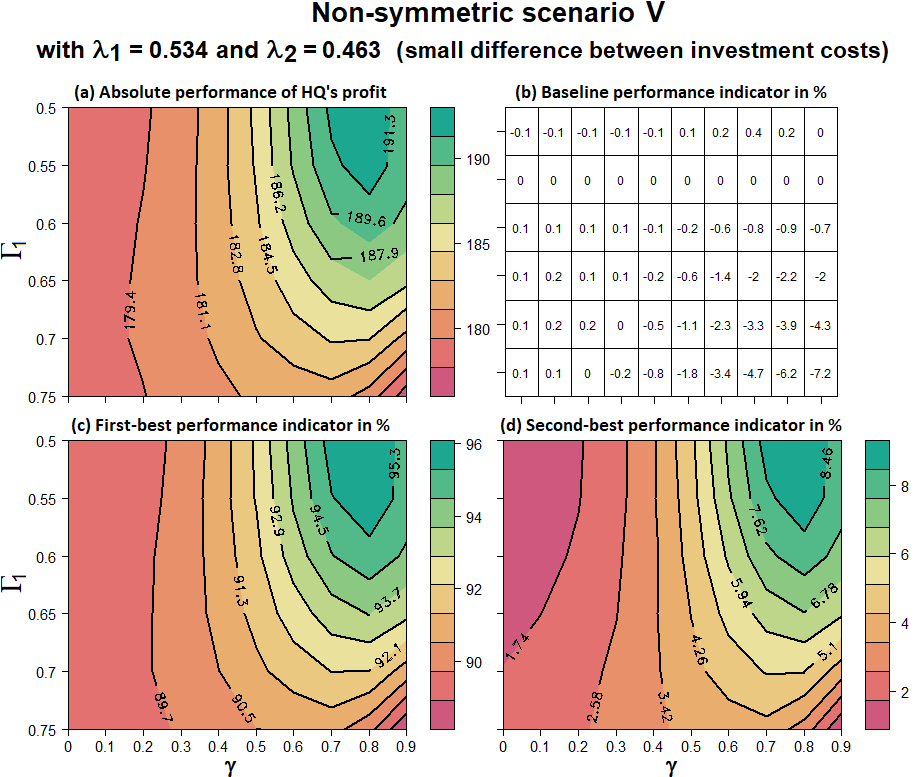}
			\caption{Results for non-symmetric marginal cost parameters with $\sigma=0$ and exploration policy is BM: (a) absolute performance of the headquarters' profit resulting from the simulation, (b) relative performance change and statistical hypothesis testing of $\Pi_{HQ}$ compared to the simulation output from the baseline scenario with $\Gamma_1^{sb} \hspace{-0.75mm} =0.55$, (c) relative performance of $\Pi_{HQ}$ compared to $\Pi_{HQ}^*$, and (d) relative performance change of $\Pi_{HQ}$ compared to $\Pi_{HQ}^{sb}$.} 
			\label{fig:Plot_35}
		\end{figure}

		\subsubsection{Scenario V}
		
\begin{figure}[b!]
			\centering
			\includegraphics[width=1\textwidth]{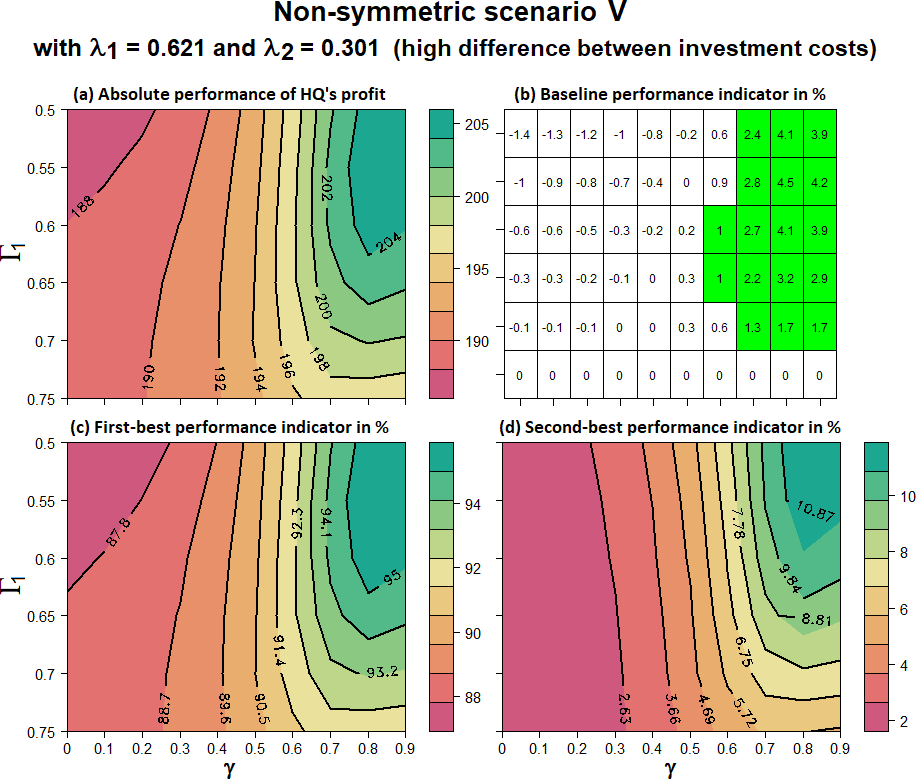}
			\caption{Results for non-symmetric marginal cost parameters with $\sigma=0$ and exploration policy is BM: (a) absolute performance of the headquarters' profit resulting from the simulation, (b) relative performance change and statistical hypothesis testing of $\Pi_{HQ}$ compared to the simulation output from the baseline scenario with $\Gamma_1^{sb} \hspace{-0.75mm} =0.75$, (c) relative performance of $\Pi_{HQ}$ compared to $\Pi_{HQ}^*$, and (d) relative performance change of $\Pi_{HQ}$ compared to $\Pi_{HQ}^{sb}$.}
			\label{fig:Plot_36}
		\end{figure}

		
			Due to symmetry considerations, it is sufficient to vary $\Gamma_1$ from $0.5$ to $0.75$ in order to study the agents' decision-making behavior in case of non-symmetric marginal cost parameters (cf. Tab. \ref{tab:Parameter_settings_and_scenario_overview}). 
			Note that, in scenarios I-IV, $\Gamma_1=\Gamma_2$, while, in scenarios V-VI, $\Gamma_2=1-\Gamma_1$. 
			For $\lambda_1=0.534$ and $\lambda_2=0.463$ (small difference between investment costs), the simulation results on the performance of fuzzy Q-learning agents are summarized in Fig. \ref{fig:Plot_35}. 
			For $\lambda_1=0.621$ and $\lambda_2=0.301$ (large difference between investment costs), the simulation outputs are depicted in Fig. \ref{fig:Plot_36}. 
			
			If the marginal cost parameters differ only slightly (cf. Fig. \ref{fig:Plot_35}), a deviation from the theory of subgame perfection equilibrium leads to a lower investment level and, thus, lower profits are made. 
			As discussed in the symmetric case, the headquarters should set the surplus sharing parameters $\Gamma_1$ and $\Gamma_2$ according to Eq. \ref{eq:Gamma_1_sb} and Eq. \ref{eq:Gamma_2_sb}, respectively, in order to maximize profits. 
			Furthermore, the simulation results show that the higher the agents' level of foresight, the higher the generated profits. 
			Moreover, the headquarters' profit as well as the divisions' investment decisions are almost normally distributed in all parameter constellations investigated. 
			
			According to Fig. \ref{fig:Plot_36}, a deviation from $\Gamma_1^{sb}$ can actually lead to higher profits, especially, if divisions have a high degree of foresight. 
			The simulation results provide support the intuition that all divisions should receive approximately the same share of the contribution margin. 
			In particular, in situations in which the headquarters only have poor information, e.g., about the costs of manufacturing, the headquarters relies on simple methods for determining decision problems in transfer pricing. 
			According to the weighted arithmetic mean of $\Gamma_j$ ($\Gamma_1=0.586$ and $\Gamma_2=0.414$), a $3:2$ surplus sharing rule for seller and first buyer and a distribution ratio of the bargaining power at a ratio of $2:3$ for seller and second buyer may be good choices for determining the surplus sharing parameters when divisions' marginal cost parameters differ widely. 

		\subsubsection{Scenario VI}
		
			As in scenario IV, an extensive sensitivity analysis is conducted to check whether the simulation results for non-symmetric marginal cost parameters are robust in terms of volatility on the markets and the implementation of other exploration policies. 
			According to figures in Appendix C, the findings appear to be robust against the turbulence of the market environment and different exploration policies.  
			
\begin{figure}[h!]
		\centering
		\includegraphics[width=0.96\textwidth]{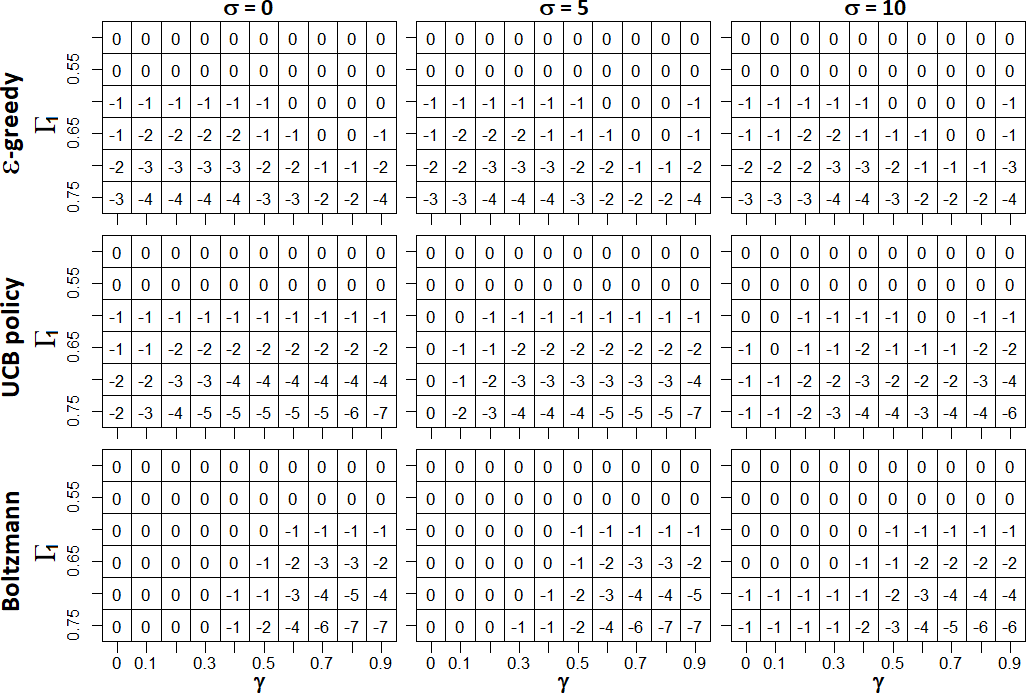}
		\caption{Results of the baseline performance indicator for $\Gamma_1^{sb}=0.5$ of different exploration policies and standard deviations of state variables given $\lambda_S=1$,  $\lambda_1=0.5$, and $\lambda_2=0.5$.}
		\label{fig:Plot_20}
	\end{figure}

\begin{figure}[h!]
		\centering
		\includegraphics[width=0.96\textwidth]{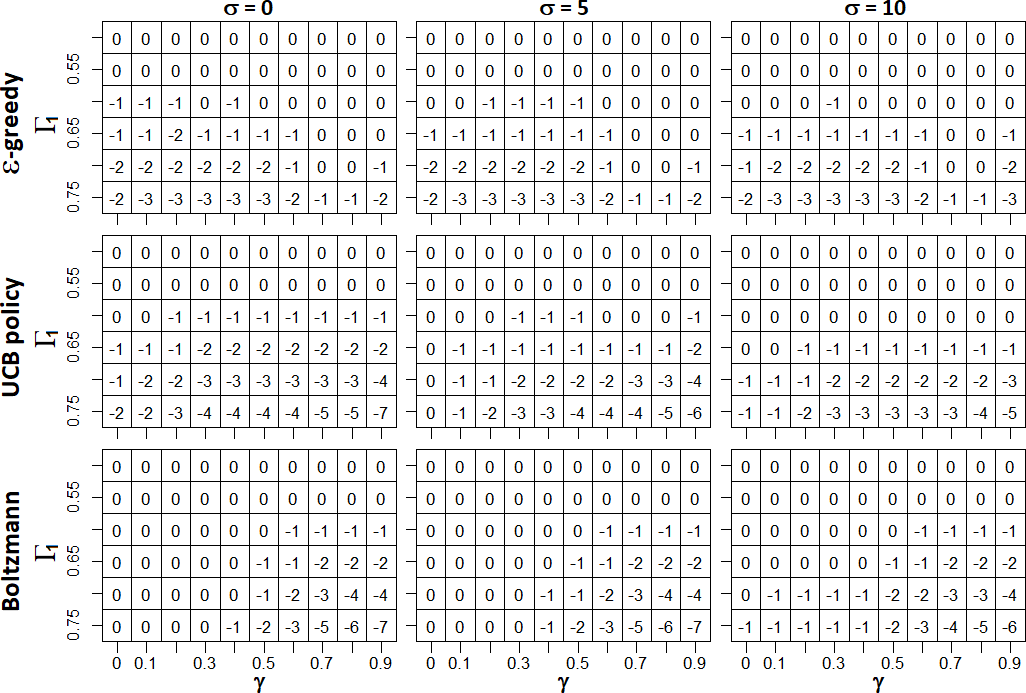}
		\caption{Results of the baseline performance indicator for $\Gamma_1^{sb}=0.55$ of different exploration policies and standard deviations of state variables given $\lambda_S=1$,  $\lambda_1=0.534$, and $\lambda_2=0.463$.}
		\label{fig:Plot_22}
	\end{figure}

\begin{figure}[h!]
		\centering
		\includegraphics[width=0.96\textwidth]{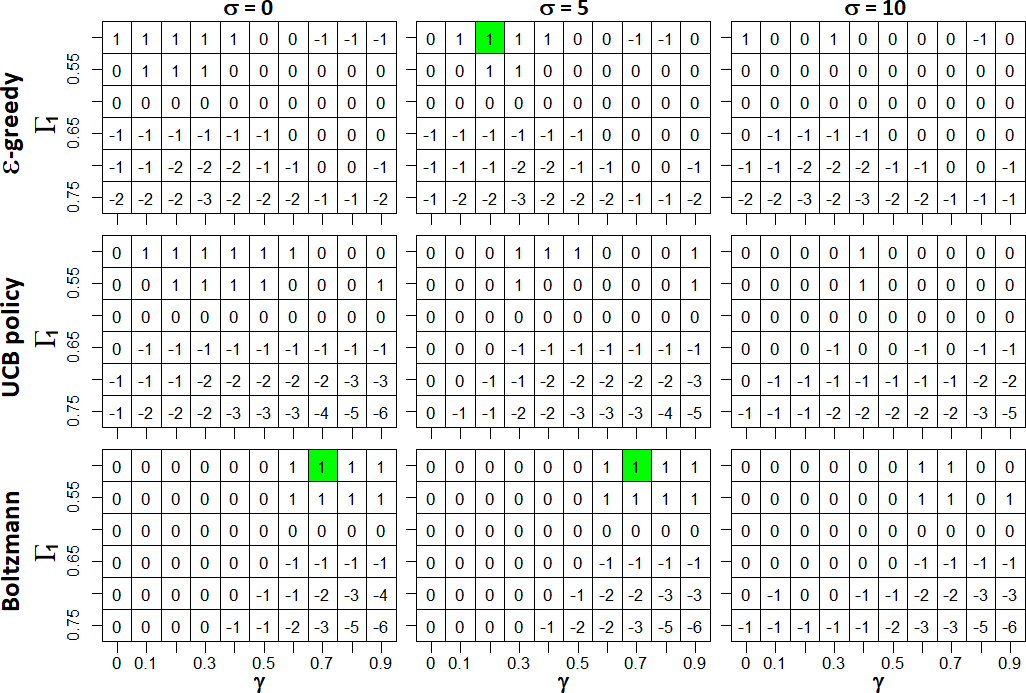}
		\caption{Results of the baseline performance indicator for $\Gamma_1^{sb}=0.6$ of different exploration policies and standard deviations of state variables given $\lambda_S=1$,  $\lambda_1=0.564$, and $\lambda_2=0.425$.}
		\label{fig:Plot_24}
	\end{figure}

\begin{figure}[h!]
		\centering
		\includegraphics[width=0.96\textwidth]{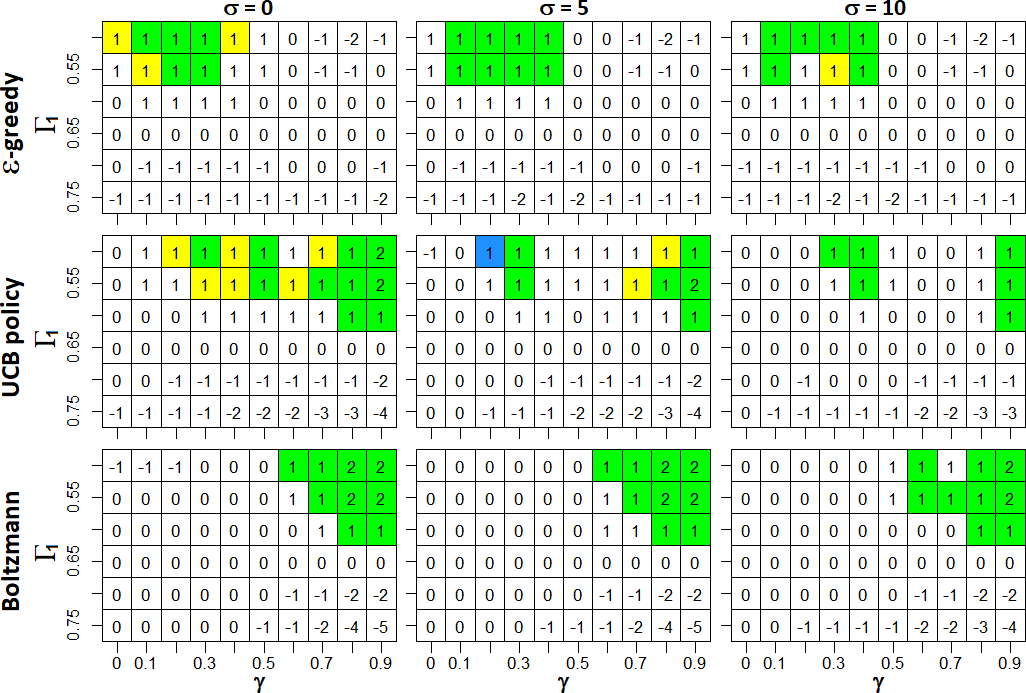}
		\caption{Results of the baseline performance indicator for $\Gamma_1^{sb}=0.65$ of different exploration policies and standard deviations of state variables given $\lambda_S=1$,  $\lambda_1=0.59$, and $\lambda_2=0.385$.}
		\label{fig:Plot_26}
	\end{figure}

\begin{figure}[h!]
		\centering
		\includegraphics[width=0.96\textwidth]{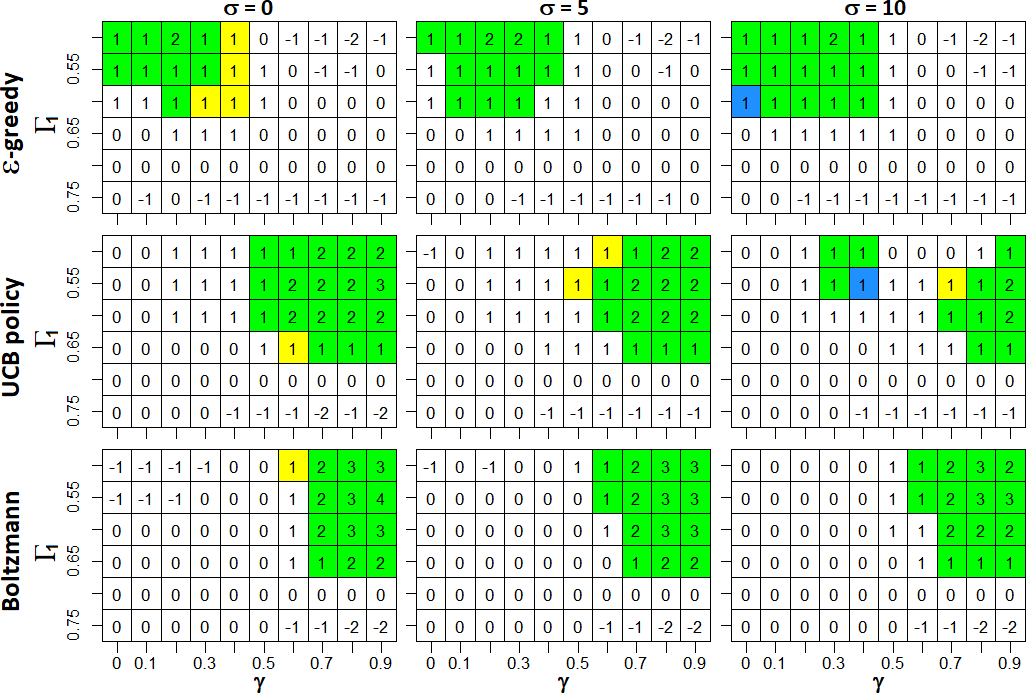}
		\caption{Results of the baseline performance indicator for $\Gamma_1^{sb}=0.7$ of different exploration policies and standard deviations of state variables given $\lambda_S=1$,  $\lambda_1=0.609$, and $\lambda_2=0.343$.}
		\label{fig:Plot_28}
	\end{figure}

\begin{figure}[h!]
		\centering
		\includegraphics[width=0.96\textwidth]{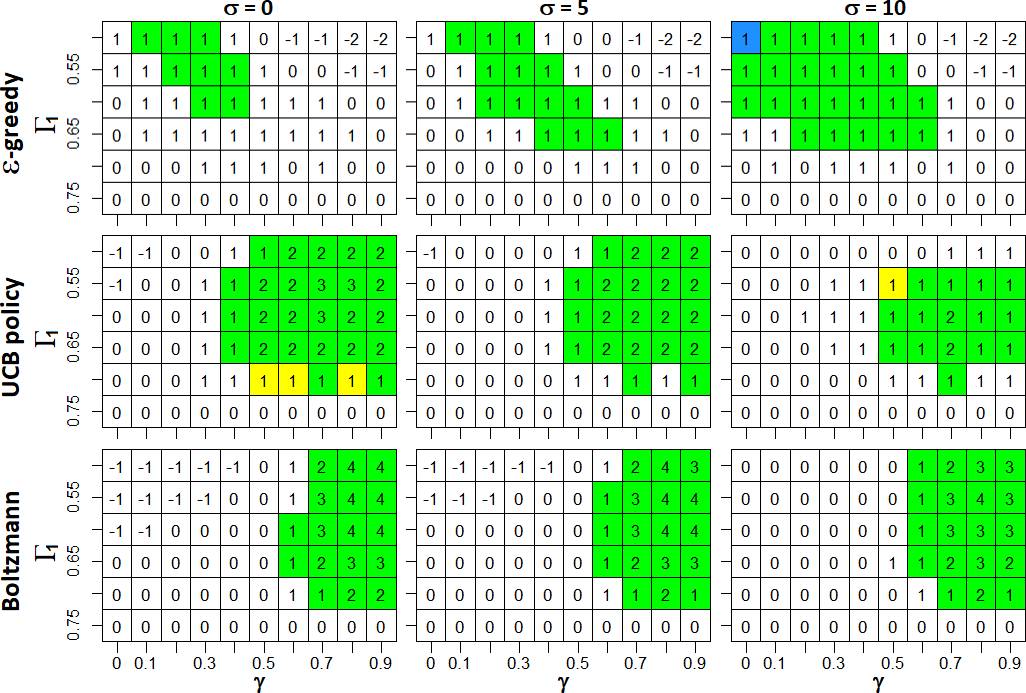}
		\caption{Results of the baseline performance indicator for $\Gamma_1^{sb}=0.75$ of different exploration policies and standard deviations of state variables given $\lambda_S=1$,  $\lambda_1=0.621$, and $\lambda_2=0.301$.}
		\label{fig:Plot_30}
	\end{figure}

			The simulation results of the baseline performance indicator in Fig. \ref{fig:Plot_20}-\ref{fig:Plot_30} suggest that, for non-myopic fuzzy Q-learning agents, there is an advantage in allocating equal bargaining power to all three divisions when marginal costs differ.  
			Adjusting the distribution ratio of bargaining power in direction of an equal level leads to higher investments and, thus, to a significant improvement in agents' performance.  

\section{Summary and conclusive remarks}
\label{sec:Conclusion}


	The starting point of the simulation study is the well-known negotiated transfer pricing model by \cite{edlin1995specific}. 
	The authors extend the neoclassical model of \cite{schmalenbach1908verrechnungspreise} and \cite{williamson1985economic} by assuming that each division can simultaneously make an upfront specific investment that enhances the value of internal trade. 
	However, in negotiated transfer pricing models with specific investments \citep[e.g.,][]{eccles1988price, edlin1995specific, vaysman1998model}, it is generally supposed that divisions operate like fully individual rational utility maximizers. 
	But in reality, such assumptions are quite demanding, because, e.g., how could a division expect that the other division will behave optimally at a later point in time as required by (Nash) equilibrium concept. 
	For this purpose, an agent-based variant of negotiated transfer pricing with individual utility maximizers who are subject to limitations is examined. 

	To deal with the divisions' bounded rationality, asymmetric information, and cognitive limitations, fuzzy Q-learning is apply. 
	Moreover, divisions are not able to find the best investment decision instantaneously, instead they have the ability to explore their decision space stepwise in order to find good policies. 
	In addition, the study widely relaxes the common knowledge assumption and also increases the complexity of the divisions' coordination problem by increasing the number of downstream divisions by one. 
		
	The paper makes two main contributions: on the one hand, the paper derives closed-form expressions for the subgame perfect equilibrium of the investment hold-up problem with one supplying division and two buying divisions and, on the other hand, the computer simulation provides some important insights into the dynamics of negotiated transfer pricing. 
	According to the simulation outputs, fuzzy Q-learning agents perform at least as well or better than fully individual rational utility maximizers. 
	In addition, the simulation results show that, if the headquarters applies the concept of subgame perfect equilibrium and allocates the distribution ratio of the bargaining power according to optimal (Nash) solutions, then fuzzy Q-learning agents generate profits that are at least in the range of the second-best solution or even higher (however, first-best solutions are generally not achieved). 
	
	In cases where both buying divisions have marginal cost parameters that are only half as large as the supplying division, the headquarters is well advised to give all of its divisions the same bargaining power. 
	But if the marginal cost parameters of both buying divisions are less than half, then the headquarters should assign the supplying division a greater bargaining power than the theory of subgame perfection equilibrium recommends. 
	In such a case, an $1:2$ surplus sharing rule for seller and buyer may be a good choice for non-myopic divisions. 
	This finding is especially important when the headquarters cannot optimally set the bargaining power due to a lack of information. 
	On the other hand, if the marginal cost parameters of the buying divisions differ widely from each other, then $2:3$ surplus sharing rules may favor divisions' investment decisions and, thus, can lead to significantly higher profits.
	The findings also show that the performance of fuzzy Q-learning agents depends crucially on the level of foresight. 
	According to an extensive sensitivity analysis, the simulation outputs appear to be robust against the turbulence of the market environment and different exploration policies. 

	Future research could address the following points: (1) The simulation study investigated here focuses on fuzzy Q-learning. 
	It would be interesting to know to what extent the simulation results change, when other reinforcement learning approaches are applied instead. 
	(2) The expansion is not limited to two buying divisions. 
	Besides, it is also interesting to examine other firm settings, e.g, a supply chain in which each division can improve the quality of the intermediate product by upfront investments. 
	(3) Future research could also study negotiated transfer pricing under capacity constraints. 
	This is especially important when the number of divisions increases. 
	For example, \cite{gavious1999transfer} and \cite{wielenberg2000negotiated} examine a two-divisional firm where the term capacity describes the level of production that is achievable with the resources currently available and could be raised by investing specific resources such as labor, machinery, or simply by overtime. 
	Model extensions with several consecutive downstream divisions would certainly be interesting, especially how well cognitively bounded agents perform under these terms. 
	(4) More research work could also consider changes in the timeline. 
	For example, the assumption that the state variables are realized before the negotiation process takes place is only to simplify the backward induction. 
	In reality, stochastic influences also take place after the negotiation phase and should therefore be taken into consideration. 

\newpage
\section*{Appendix A: Flow diagram of the agent-based simulation }
\label{appendix_A:Flow_diagram}	
	\vspace{-1mm}
	\begin{figure}[b!]
		\centering
		\includegraphics[width=0.95\textwidth]{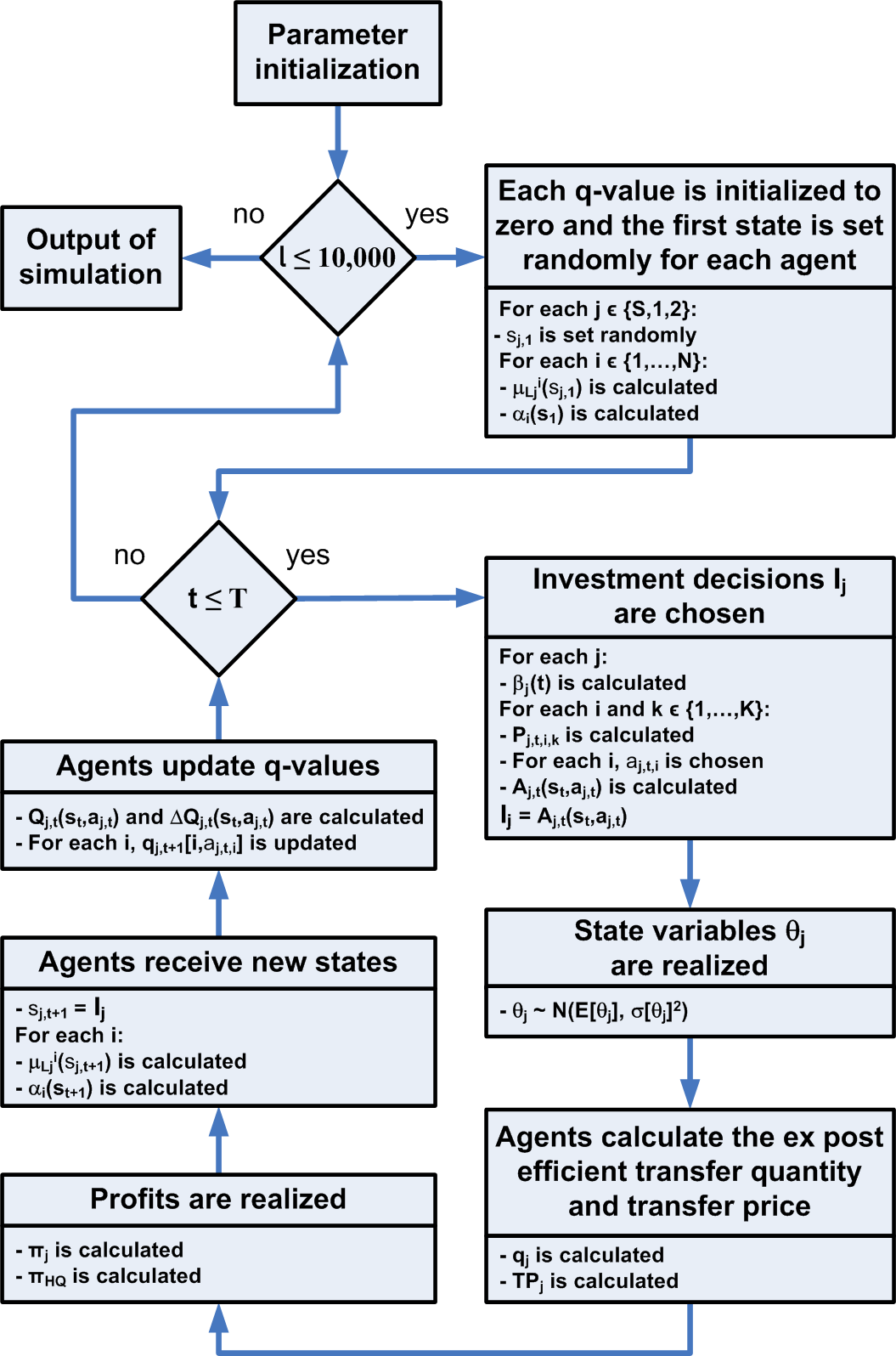}
		\caption{Flow diagram of the agent-based simulation using the Boltzmann exploration policy. For an improved readability, most for-loops are omitted. } 
		\label{fig:Procedure}
	\end{figure}

	\afterpage{
		\begin{landscape}
		\begin{table}[h!]
		\linespread{1.0}\selectfont
		\centering
		\caption{First-best and second-best solutions resulting from the concept of subgame perfect equilibrium for $b=12$, $E[\theta_S]=60$, and $E[\theta_B]=100$.}
		\label{tab:Firstbest_Secondbest_Solutions}
		\begin{tabularx}{151.5mm}{|ccccc|ccccccccc|}
			\hline
			\multicolumn{14}{|l|}{\textbf{First-best solutions for symmetric scenarios I-IV}} \\
			\hline
			$\Gamma_1$ & $\Gamma_2$ & $\lambda_S$ & $\lambda_1$ & $\lambda_2$ & $I_S^*$ & $I_1^*$ & $I_2^*$ & $q_1^*$ & $q_2^*$ & $\Pi_S^*$ & $\Pi_1^*$ & $\Pi_2^*$ & $\Pi_{HQ}^*$ \\
			\hline			
			0.5 & 0.5 & 1 & 0.5 & 0.5 & 10 & 10 & 10 & 5 & 5 & 100 & 50 & 50 & 200 \\
			0.45 & 0.45 & 1 & 0.424 & 0.424 & 10.47 & 12.35 & 12.35 & 5.23 & 5.23 & 93.10 & 58.14 & 58.14 & 209.38 \\
			0.40 & 0.40 & 1 & 0.361 & 0.361 & 11.07 & 15.33 & 15.33 & 5.53 & 5.53 & 85.67 & 67.82 & 67.82 & 221.30 \\
			0.35 & 0.35 & 1 & 0.307 & 0.307 & 11.86 & 19.32 & 19.32 & 5.93 & 5.93 & 77.13 & 80.08 & 80.08 & 237.29 \\
			0.30 & 0.30 & 1 & 0.262 & 0.262 & 12.94 & 24.69 & 24.69 & 6.47 & 6.47 & 67.02 & 95.87 & 95.87 & 258.77 \\
			0.25 & 0.25 & 1 & 0.222 & 0.222 & 14.56 & 32.79 & 32.79 & 7.28 & 7.28 & 52.79 & 119.18 & 119.18 & 291.15 \\
			\hline
			\multicolumn{14}{|l|}{\textbf{Second-best solutions for symmetric scenarios I-IV}} \\
			\hline
			$\Gamma_1$ & $\Gamma_2$ & $\lambda_S$ & $\lambda_1$ & $\lambda_2$ & $I_S^{sb}$ & $I_1^{sb}$ & $I_2^{sb}$ & $q_1^{sb}$ & $q_2^{sb}$ & $\Pi_S^{sb}$ & $\Pi_1^{sb}$ & $\Pi_2^{sb}$ & $\Pi_{HQ}^{sb}$ \\
			\hline			
			0.5 & 0.5 & 1 & 0.5 & 0.5 & 4 & 4 & 4 & 4 & 4 & 88 & 44 & 44 & 176 \\
			0.45 & 0.45 & 1 & 0.424 & 0.424 & 3.67 & 5.29 & 5.29 & 4.08 & 4.08 & 83.14 & 49.02 & 49.02 & 181.18 \\
			0.40 & 0.40 & 1 & 0.361 & 0.361 & 3.35 & 6.97 & 6.97 & 4.19 & 4.19 & 78.78 & 54.55 & 54.55 & 187.89 \\
			0.35 & 0.35 & 1 & 0.307 & 0.307 & 3.04 & 9.23 & 9.23 & 4.36 & 4.36 & 74.92 & 61.01 & 61.01 & 196.93 \\
			0.30 & 0.30 & 1 & 0.262 & 0.262 & 2.75 & 12.24 & 12.24 & 4.58 & 4.58 & 71.85 & 68.56 & 68.56 & 208.97 \\
			0.25 & 0.25 & 1 & 0.222 & 0.222 & 2.46 & 16.65 & 16.65 & 4.93 & 4.93 & 69.67 & 78.46 & 78.46 & 226.59 \\ 
			\hline
			\noalign{\vskip 1.8mm}   
			\hline 
			\multicolumn{14}{|l|}{\textbf{First-best solutions for non-symmetric scenarios V-VI}} \\
			\hline
			$\Gamma_1$ & $\Gamma_2$ & $\lambda_S$ & $\lambda_1$ & $\lambda_2$ & $I_S^*$ & $I_1^*$ & $I_2^*$ & $q_1^*$ & $q_2^*$ & $\Pi_S^*$ & $\Pi_1^*$ & $\Pi_2^*$ & $\Pi_{HQ}^*$ \\
			\hline			
			0.5 & 0.5 & 1 & 0.5 & 0.5 & 10 & 10 & 10 & 5 & 5 & 100 & 50 & 50 & 200 \\
			0.55 & 0.45 & 1 & 0.534 & 0.463 & 10.02 & 9.25 & 10.98 & 4.94 & 5.08 & 100.02 & 42.96 & 57.48 & 200.46 \\
			0.60 & 0.40 & 1 & 0.564 & 0.425 & 10.09 & 8.68 & 12.22 & 4.90 & 5.19 & 100.12 & 36.32 & 65.36 & 201.80 \\
			0.65 & 0.35 & 1 & 0.590 & 0.385 & 10.21 & 8.26 & 13.87 & 4.87 & 5.34 & 100.32 & 29.73 & 74.21 & 204.26 \\
			0.70 & 0.30 & 1 & 0.609 & 0.343 & 10.42 & 7.99 & 16.18 & 4.87 & 5.55 & 100.56 & 23.18 & 84.61 & 208.35 \\
			0.75 & 0.25 & 1 & 0.621 & 0.301 & 10.73 & 7.86 & 19.42 & 4.88 & 5.85 & 101.05 & 16.36 & 97.16 & 214.57 \\
			\hline
			\multicolumn{14}{|l|}{\textbf{Second-best solutions for non-symmetric scenarios V-VI}} \\
			\hline
			$\Gamma_1$ & $\Gamma_2$ & $\lambda_S$ & $\lambda_1$ & $\lambda_2$ & $I_S^{sb}$ & $I_1^{sb}$ & $I_2^{sb}$ & $q_1^{sb}$ & $q_2^{sb}$ & $\Pi_S^{sb}$ & $\Pi_1^{sb}$ & $\Pi_2^{sb}$ & $\Pi_{HQ}^{sb}$ \\
			\hline			
			0.5 & 0.5 & 1 & 0.5 & 0.5 & 4 & 4 & 4 & 4 & 4 & 88 & 44 & 44 & 176 \\
			0.55 & 0.45 & 1 & 0.534 & 0.463 & 4.00 & 3.32 & 4.84 & 3.94 & 4.07 & 88.02 & 39.00 & 49.31 & 176.32 \\
			0.60 & 0.40 & 1 & 0.564 & 0.425 & 4.00 & 2.76 & 5.87 & 3.90 & 4.16 & 88.10 & 34.30 & 54.86 & 177.26 \\
			0.65 & 0.35 & 1 & 0.590 & 0.385 & 4.00 & 2.29 & 7.21 & 3.86 & 4.27 & 88.26 & 29.70 & 61.03 & 178.99 \\
			0.70 & 0.30 & 1 & 0.609 & 0.343 & 4.00 & 1.88 & 9.03 & 3.82 & 4.42 & 88.47 & 25.23 & 68.13 & 181.82 \\
			0.75 & 0.25 & 1 & 0.621 & 0.301 & 4.00 & 1.52 & 11.55 & 3.79 & 4.63 & 88.92 & 20.75 & 76.46 & 186.13 \\
			\hline
		\end{tabularx}
		\linespread{1}\selectfont
		\end{table}
		\end{landscape}
	}

\newpage
\section*{Appendix B: Sensitivity analysis conducted according to scenario IV}
\label{appendix_B:Sensitivity_analysis}
	\vspace{-5mm}
	\begin{figure}[h!]
		\centering
		\includegraphics[width=0.99\textwidth]{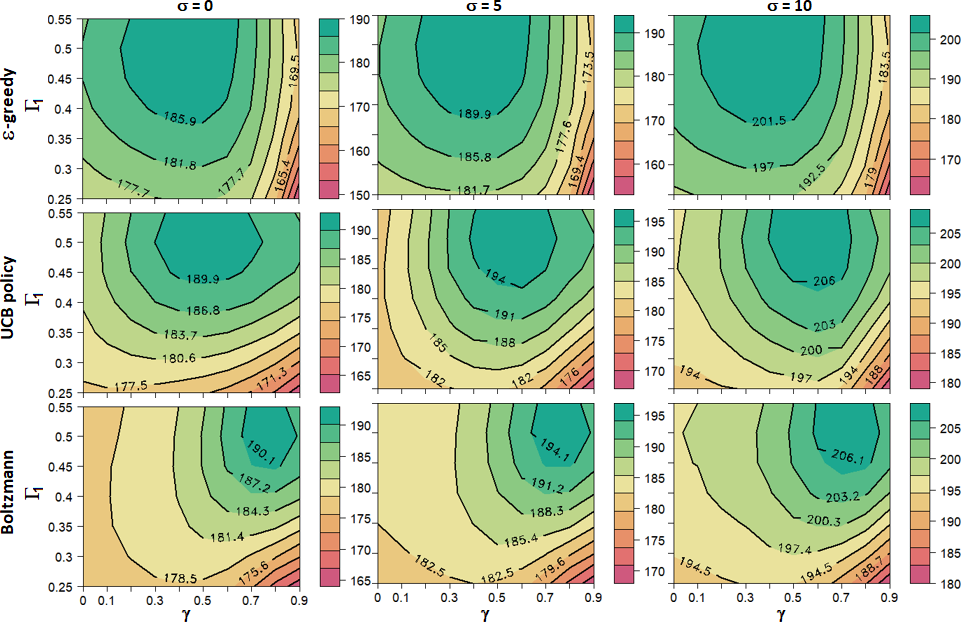}
		\caption{Abs. performance of the headquarters' profit for $\Gamma_1^{sb}=0.5$ of different exploration policies and standard deviations of state variables given $\lambda_S=1$ and $\lambda_1=\lambda_2=0.5$.}
		\label{fig:Plot_07}
	\end{figure}
	\vspace{-5mm}
	\begin{figure}[h!]
		\centering
		\includegraphics[width=0.99\textwidth]{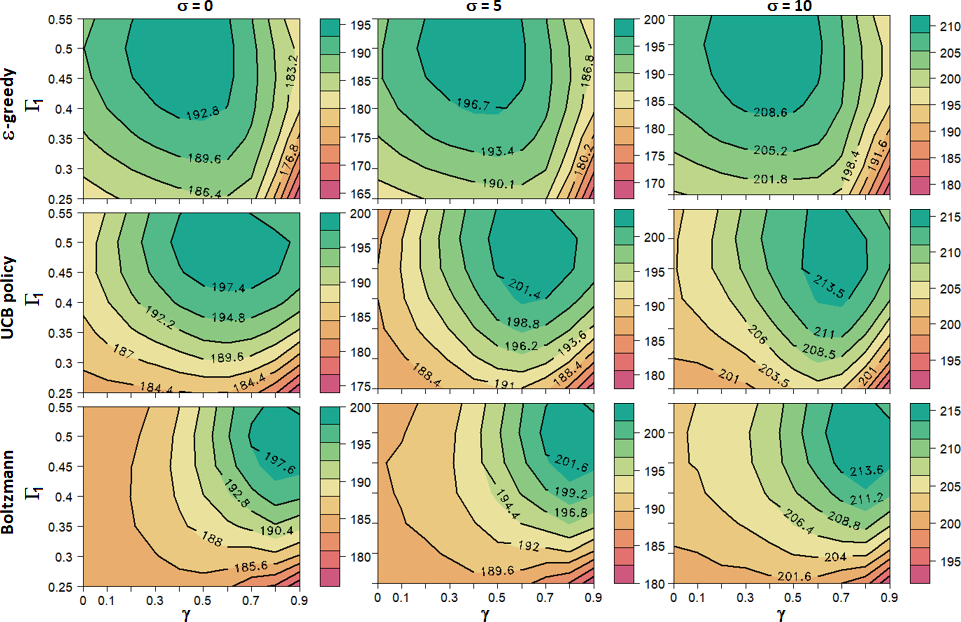}
		\caption{Abs. performance of the headquarters' profit for $\Gamma_1^{sb}=0.45$ of different exploration policies and standard deviations of state variables given $\lambda_S=1$ and $\lambda_1=\lambda_2=0.424$.}
		\label{fig:Plot_09}
	\end{figure}

\newpage	
	\begin{figure}[h!]
		\centering
		\includegraphics[width=0.99\textwidth]{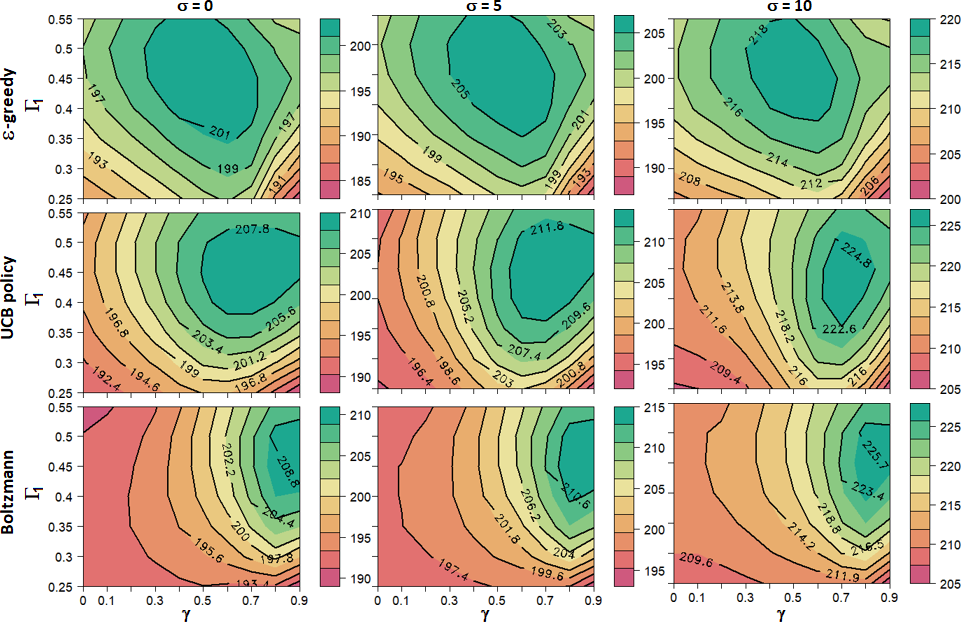}
		\caption{Abs. performance of the headquarters' profit for $\Gamma_1^{sb}=0.4$ of different exploration policies and standard deviations of state variables given $\lambda_S=1$ and $\lambda_1=\lambda_2=0.361$.}
		\label{fig:Plot_11}
	\end{figure}
	\begin{figure}[h!]
		\centering
		\includegraphics[width=0.99\textwidth]{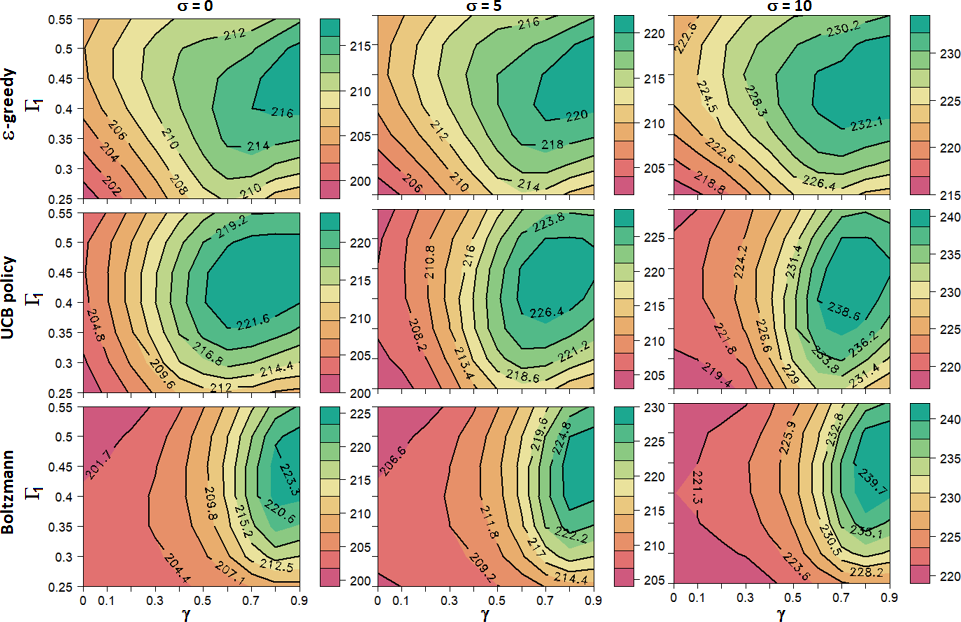}
		\caption{Abs. performance of the headquarters' profit for $\Gamma_1^{sb}=0.35$ of different exploration policies and standard deviations of state variables given $\lambda_S=1$ and $\lambda_1=\lambda_2=0.307$.}
		\label{fig:Plot_13}
	\end{figure}

\newpage
	\begin{figure}[h!]
		\centering
		\includegraphics[width=0.99\textwidth]{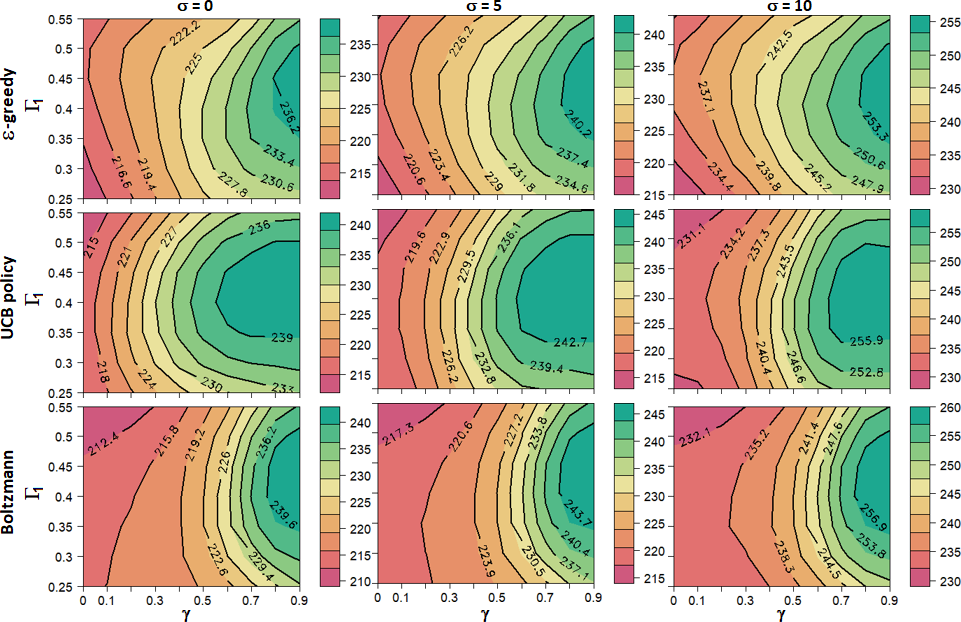}
		\caption{Abs. performance of the headquarters' profit for $\Gamma_1^{sb}=0.3$ of different exploration policies and standard deviations of state variables given $\lambda_S=1$ and $\lambda_1=\lambda_2=0.262$.}
		\label{fig:Plot_15}
	\end{figure}
	\begin{figure}[h!]
		\centering
		\includegraphics[width=0.95\textwidth]{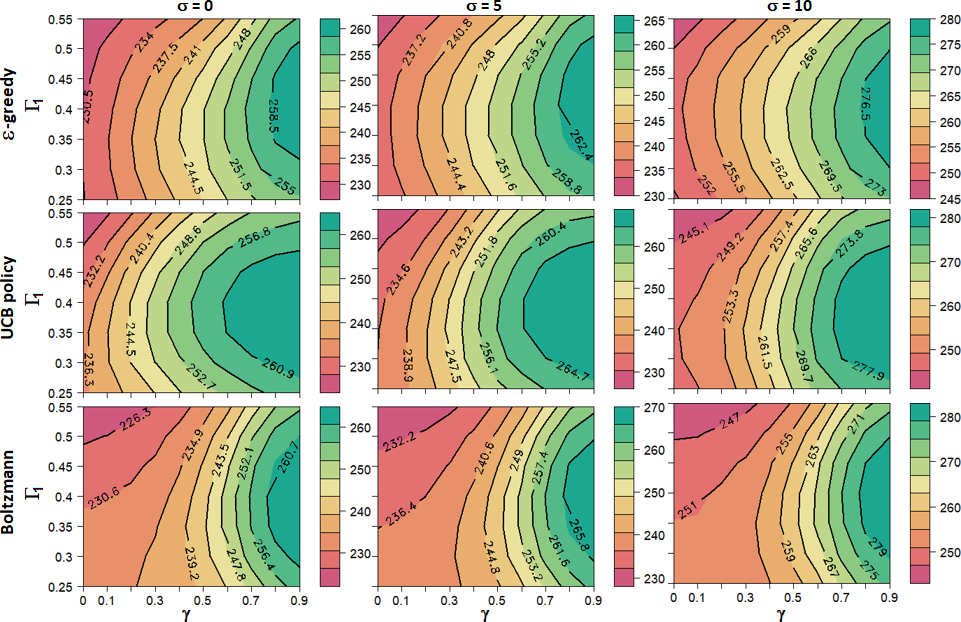}
		\caption{Abs. performance of the headquarters' profit for $\Gamma_1^{sb}=0.25$ of different exploration policies and standard deviations of state variables given $\lambda_S=1$ and $\lambda_1=\lambda_2=0.222$.}
		\label{fig:Plot_17}
	\end{figure}

\newpage	
	\begin{figure}[h!]
		\centering
		\includegraphics[width=0.96\textwidth]{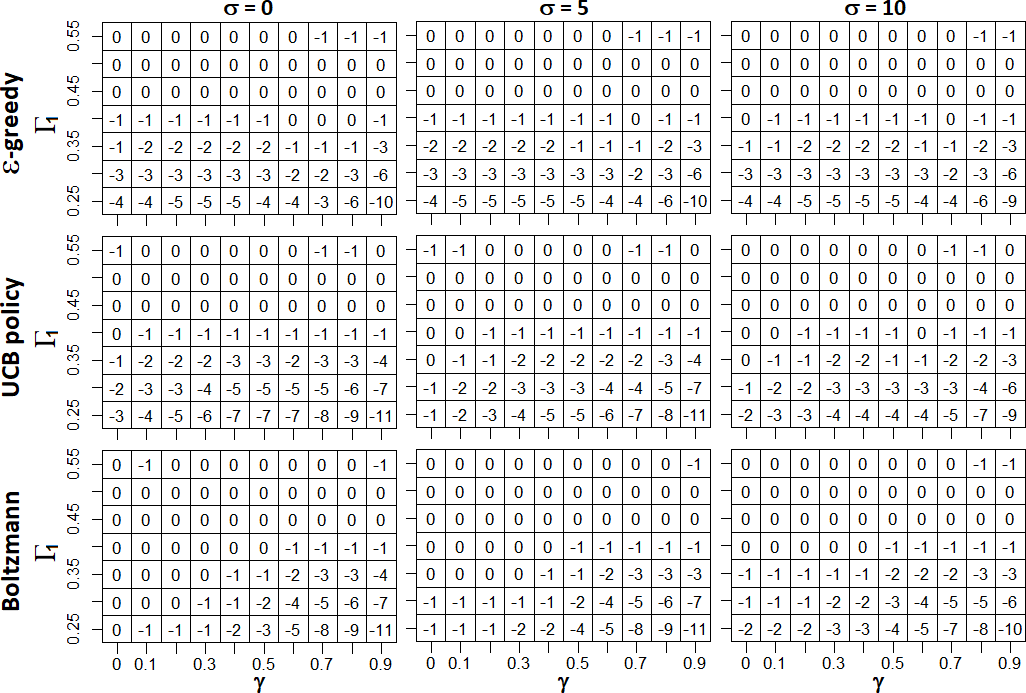}
		\caption{Results of the baseline performance indicator for $\Gamma_j^{sb}=0.45$ of different exploration policies and standard deviations of state variables given $\lambda_S=1$ and $\lambda_1=\lambda_2=0.424$.}
		\label{fig:Plot_10}
	\end{figure}

\newpage	
	\begin{figure}[h!]
		\centering
		\includegraphics[width=0.96\textwidth]{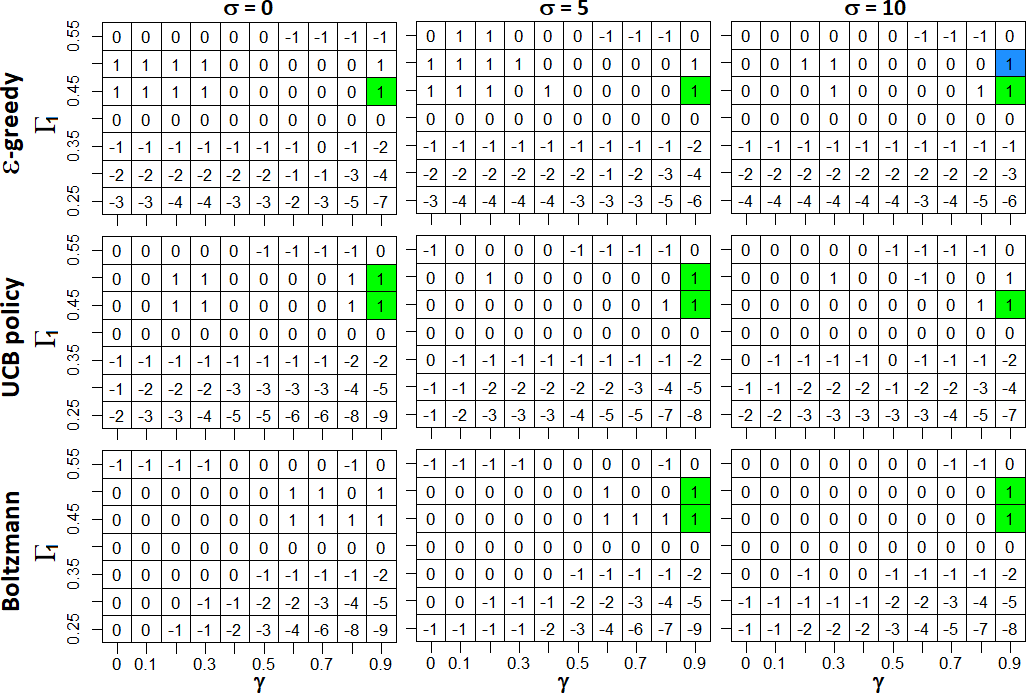}
		\caption{Results of the baseline performance indicator for $\Gamma_j^{sb}=0.4$ of different exploration policies and standard deviations of state variables given $\lambda_S=1$ and $\lambda_1=\lambda_2=0.361$.}
		\label{fig:Plot_12}
	\end{figure}	
	\begin{figure}[h!]
		\centering
		\includegraphics[width=0.96\textwidth]{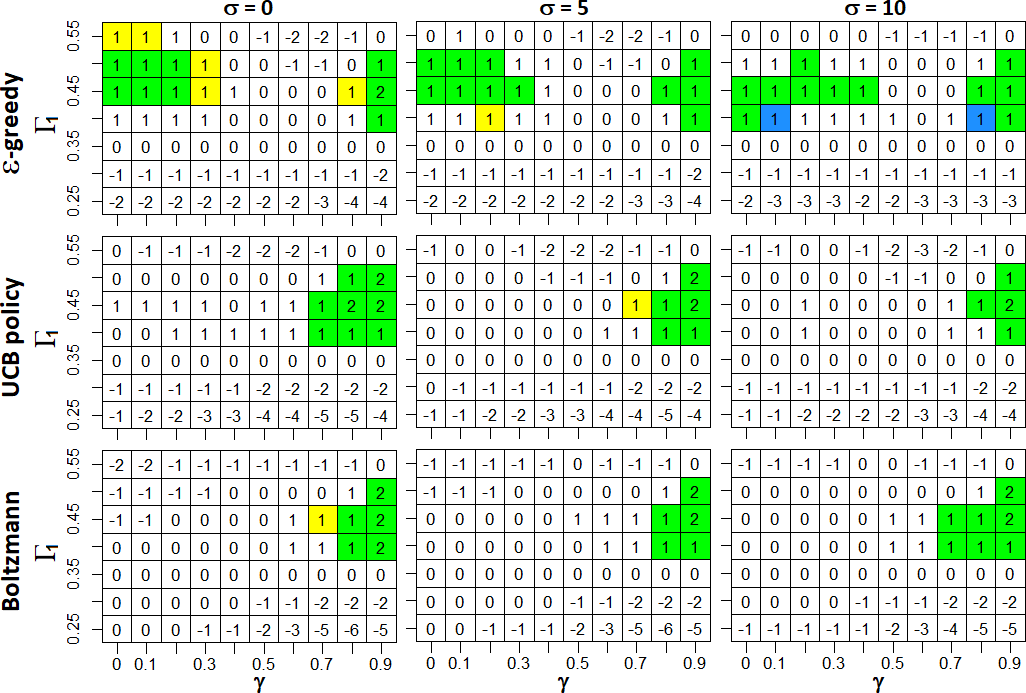}
		\caption{Results of the baseline performance indicator for $\Gamma_j^{sb}=0.35$ of different exploration policies and standard deviations of state variables given $\lambda_S=1$ and $\lambda_1=\lambda_2=0.307$.}
		\label{fig:Plot_14}
	\end{figure}

\newpage	
	\begin{figure}[h!]
		\centering
		\includegraphics[width=0.96\textwidth]{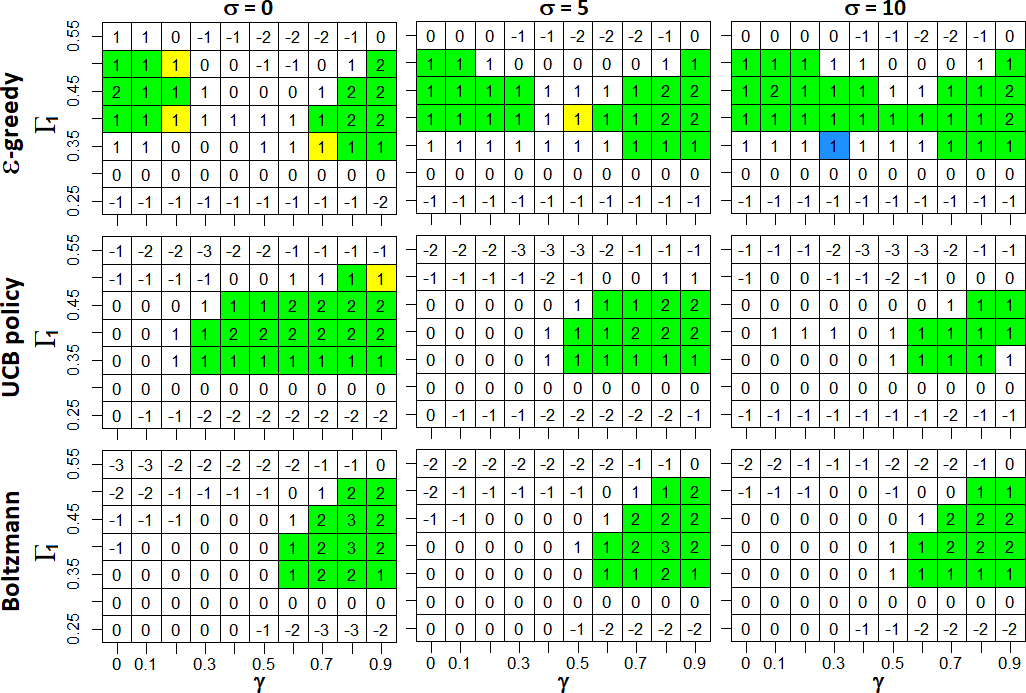}
		\caption{Results of the baseline performance indicator for $\Gamma_j^{sb}=0.3$ of different exploration policies and standard deviations of state variables given $\lambda_S=1$ and $\lambda_1=\lambda_2=0.262$.}
		\label{fig:Plot_16}
	\end{figure}	
	\begin{figure}[h!]
		\centering
		\includegraphics[width=0.96\textwidth]{Plot_18.png}
		\caption{Results of the baseline performance indicator for $\Gamma_j^{sb}=0.25$ of different exploration policies and standard deviations of state variables given $\lambda_S=1$ and $\lambda_1=\lambda_2=0.222$.}
		\label{fig:Plot_18}
	\end{figure}

\newpage
\section*{Appendix C: Sensitivity analysis conducted according to scenario VI}
\label{appendix_C:Sensitivity_analysis}
	\vspace{-5mm}
	\begin{figure}[h!]
		\centering
		\includegraphics[width=0.99\textwidth]{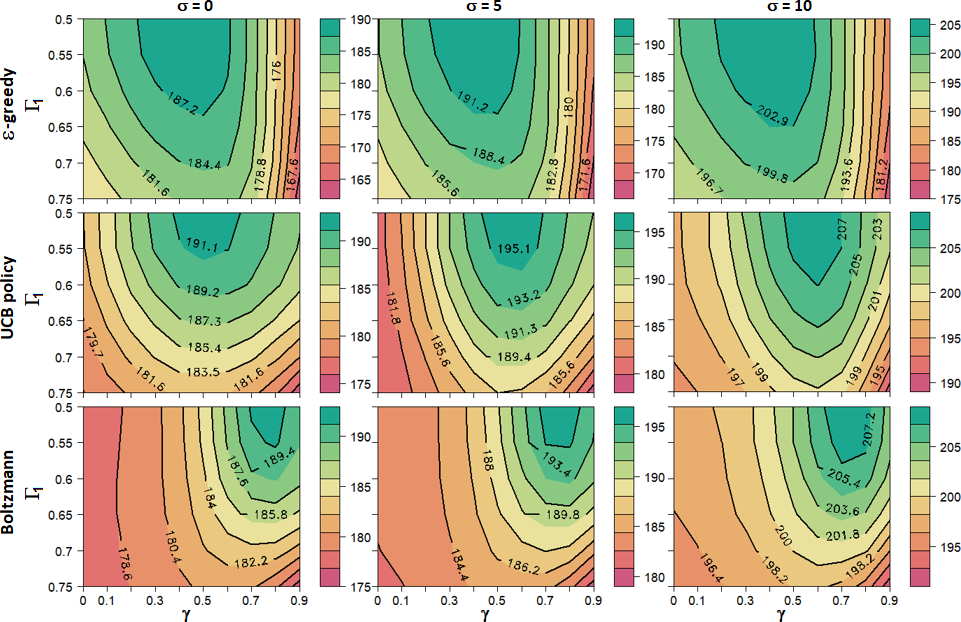}
		\caption{Abs. performance of the headquarters' profit for $\Gamma_1^{sb}=0.5$ of different exploration policies and standard deviations of state variables given $\lambda_S=1$ and $\lambda_1=\lambda_2=0.5$.}
		\label{fig:Plot_19}
	\end{figure}
	\vspace{-5mm}
	\begin{figure}[h!]
		\centering
		\includegraphics[width=0.99\textwidth]{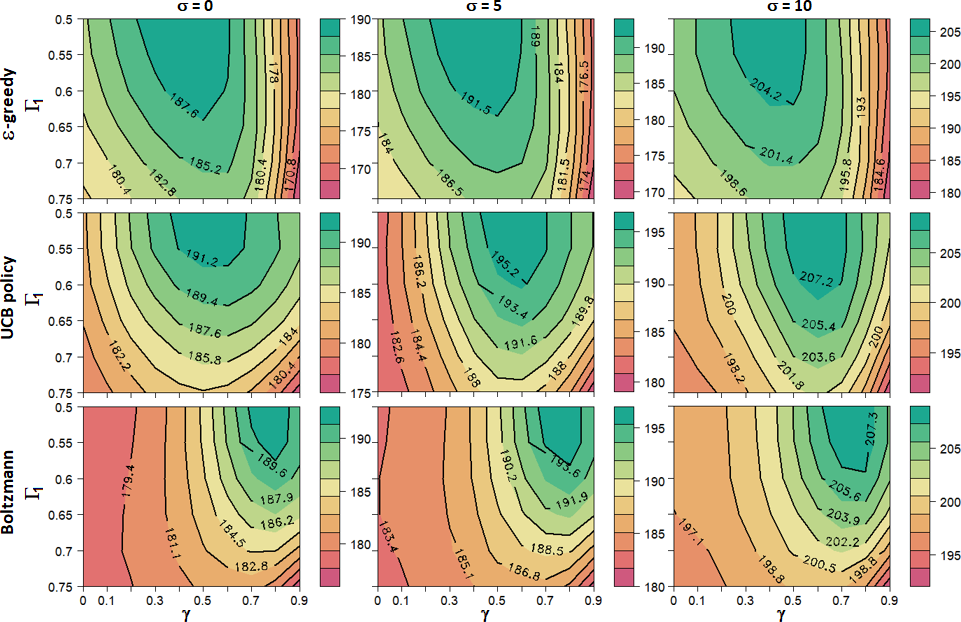}
		\caption{Abs. performance of the headquarters' profit for $\Gamma_1^{sb}=0.55$ of different exploration policies and standard deviations of state variables given $\lambda_S=1$, $\lambda_1=0.534$, and $\lambda_2=0.463$.}
		\label{fig:Plot_21}
	\end{figure}

\newpage	
	\begin{figure}[h!]
		\centering
		\includegraphics[width=0.99\textwidth]{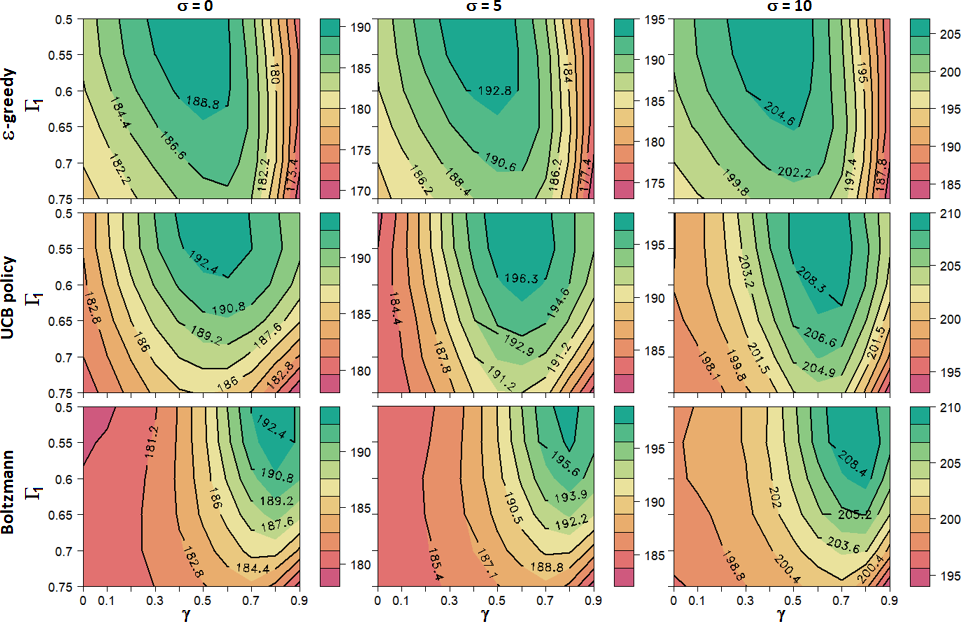}
		\caption{Abs. performance of the headquarters' profit for $\Gamma_1^{sb}=0.6$ of different exploration policies and standard deviations of state variables given $\lambda_S=1$, $\lambda_1=0.564$, and $\lambda_2=0.425$.}
		\label{fig:Plot_23}
	\end{figure}
	\begin{figure}[h!]
		\centering
		\includegraphics[width=0.99\textwidth]{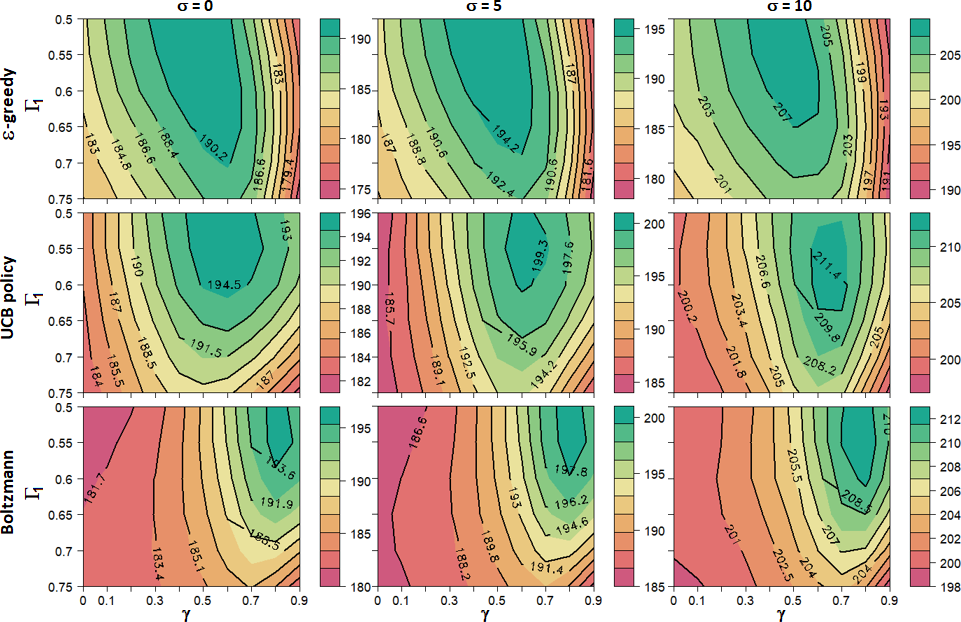}
		\caption{Abs. performance of the headquarters' profit for $\Gamma_1^{sb}=0.65$ of different exploration policies and standard deviations of state variables given $\lambda_S=1$, $\lambda_1=0.59$, and $\lambda_2=0.385$.}
		\label{fig:Plot_25}
	\end{figure}

\newpage	
	\begin{figure}[h!]
		\centering
		\includegraphics[width=0.99\textwidth]{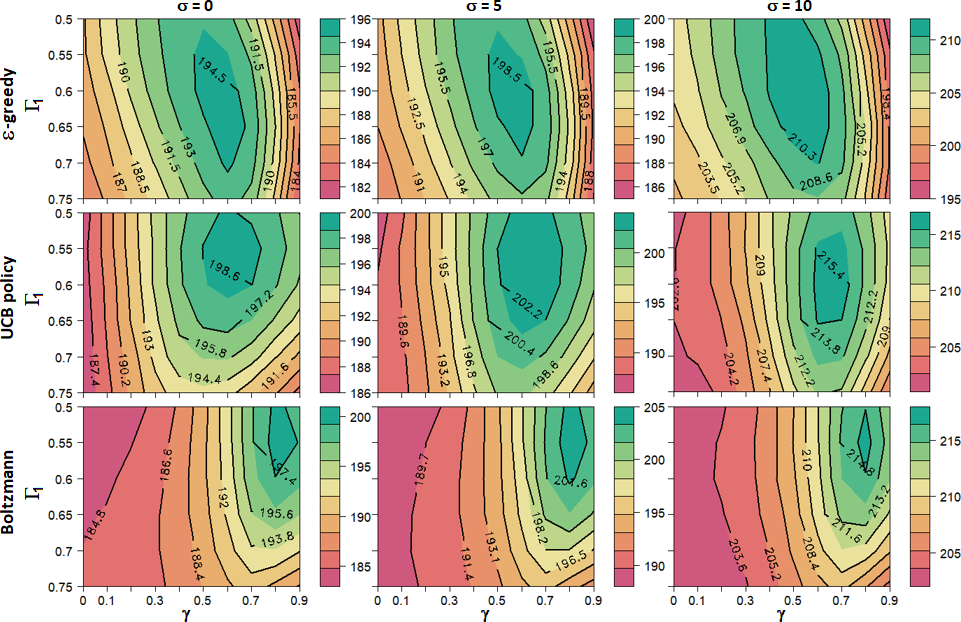}
		\caption{Abs. performance of the headquarters' profit for $\Gamma_1^{sb}=0.7$ of different exploration policies and standard deviations of state variables given $\lambda_S=1$, $\lambda_1=0.609$, and $\lambda_2=0.343$.}
		\label{fig:Plot_27}
	\end{figure}
	\begin{figure}[h!]
		\centering
		\includegraphics[width=0.99\textwidth]{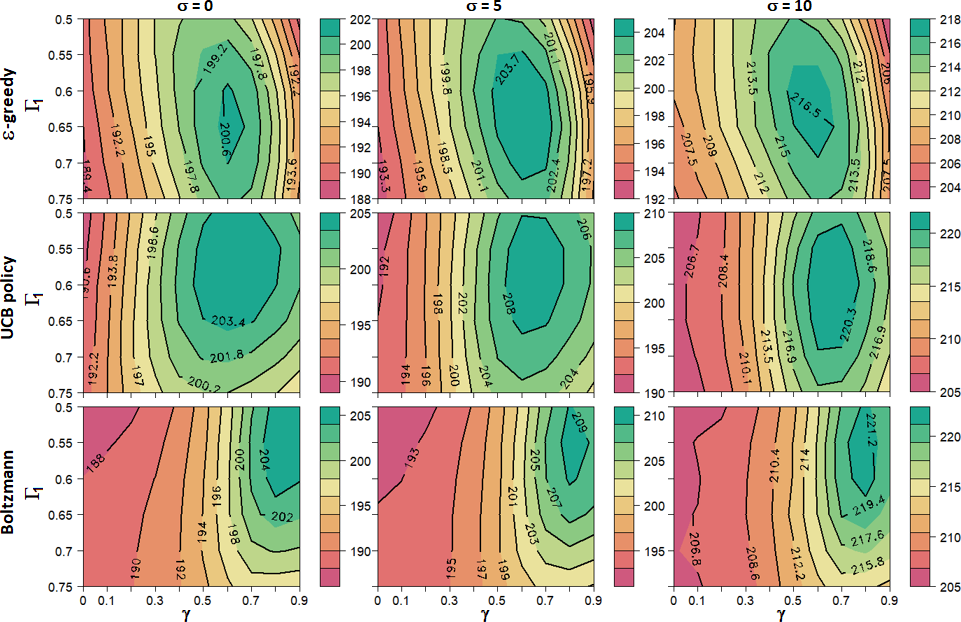}
		\caption{Abs. performance of the headquarters' profit for $\Gamma_1^{sb}=0.75$ of different exploration policies and standard deviations of state variables given $\lambda_S=1$, $\lambda_1=0.621$, and $\lambda_2=0.301$.}
		\label{fig:Plot_29}
	\end{figure}

\newpage	

\newpage	

\newpage	

\newpage
\bibliographystyle{spbasic}      
\bibliography{biblio}   

\begin{thebibliography}{36}
\providecommand{\natexlab}[1]{#1}
\providecommand{\url}[1]{{#1}}
\providecommand{\urlprefix}{URL }
\expandafter\ifx\csname urlstyle\endcsname\relax
  \providecommand{\doi}[1]{DOI~\discretionary{}{}{}#1}\else
  \providecommand{\doi}{DOI~\discretionary{}{}{}\begingroup
  \urlstyle{rm}\Url}\fi
\providecommand{\eprint}[2][]{\url{#2}}

\bibitem[{Anctil and Dutta(1999)}]{anctil1999negotiated}
Anctil RM, Dutta S (1999) Negotiated transfer pricing and divisional vs.
  firm-wide performance evaluation. The Accounting Review 74(1):87--104

\bibitem[{Axtell(2007)}]{axtell2007economic}
Axtell RL (2007) What economic agents do: How cognition and interaction lead to
  emergence and complexity. The Review of Austrian Economics 20(2):105--122

\bibitem[{Baldenius(2000)}]{baldenius2000intrafirm}
Baldenius T (2000) Intrafirm trade, bargaining power, and specific investments.
  Review of Accounting Studies 5(1):27--56

\bibitem[{Baldenius et~al.(1999)Baldenius, Reichelstein, and
  Sahay}]{baldenius1999negotiated}
Baldenius T, Reichelstein S, Sahay SA (1999) Negotiated versus cost-based
  transfer pricing. Review of Accounting Studies 4(2):67--91

\bibitem[{Cesa-Bianchi et~al.(2017)Cesa-Bianchi, Gentile, Lugosi, and
  Neu}]{cesa2017boltzmann}
Cesa-Bianchi N, Gentile C, Lugosi G, Neu G (2017) Boltzmann exploration done
  right. In: Advances in Neural Information Processing Systems, pp 6284--6293

\bibitem[{Dearden et~al.(1998)Dearden, Friedman, and
  Russell}]{dearden1998bayesian}
Dearden R, Friedman N, Russell S (1998) Bayesian q-learning. In: Proceedings of
  the Fifteenth National Conference on Artificial Intelligence, pp 761--768

\bibitem[{Eccles and White(1988)}]{eccles1988price}
Eccles RG, White HC (1988) Price and authority in inter-profit center
  transactions. American Journal of Sociology 94:17--51

\bibitem[{Edlin and Reichelstein(1995)}]{edlin1995specific}
Edlin AS, Reichelstein S (1995) Specific investment under negotiated transfer
  pricing: An efficiency result. Accounting Review 70(2):275--291

\bibitem[{Epstein(2006)}]{epstein2006generative}
Epstein JM (2006) Agent-based computational models and generative social
  science. In: Generative social science: Studies in agent-based computational
  modeling, Princeton University Press, pp 4--46

\bibitem[{Gavious(1999)}]{gavious1999transfer}
Gavious A (1999) Transfer pricing under capacity constraints. Journal of
  Accounting, Auditing \& Finance 14(1):57--72

\bibitem[{Glorennec(1994)}]{glorennec1994fuzzy}
Glorennec PY (1994) Fuzzy q-learning and dynamical fuzzy q-learning. In:
  Proceedings of 1994 IEEE 3rd International Fuzzy Systems Conference, IEEE, pp
  474--479

\bibitem[{G{\"o}x and Schiller(2006)}]{gox2006economic}
G{\"o}x RF, Schiller U (2006) An economic perspective on transfer pricing.
  Handbooks of Management Accounting Research 2:673--695

\bibitem[{Hofmann and Pfeiffer(2006)}]{hofmann2006verfugungsrechte}
Hofmann C, Pfeiffer T (2006) {Verf{\"u}gungsrechte und spezifische
  Investitionen: Steuerung {\"u}ber Budgets oder Verrechnungspreise?}
  {Schmalenbachs Zeitschrift f{\"u}r betriebswirtschaftliche Forschung}
  58(4):426--454

\bibitem[{Kofinas et~al.(2018)Kofinas, Dounis, and Vouros}]{kofinas2018fuzzy}
Kofinas P, Dounis A, Vouros G (2018) Fuzzy q-learning for multi-agent
  decentralized energy management in microgrids. Applied Energy 219:53--67

\bibitem[{Lengsfeld and Schiller(2003)}]{lengsfeld2004transfer}
Lengsfeld S, Schiller U (2003) Transfer pricing based on actual versus standard
  costs. Universit{\"a}t T{\"u}bingen T{\"u}binger Diskussionsbeitr{\"a}ge(272)

\bibitem[{Lin et~al.(2013)Lin, Lucas~Jr, and Shmueli}]{lin2013research}
Lin M, Lucas~Jr HC, Shmueli G (2013) Research commentary: too big to fail:
  large samples and the p-value problem. Information Systems Research
  24(4):906--917

\bibitem[{Mann and Whitney(1947)}]{mann1947test}
Mann HB, Whitney DR (1947) On a test of whether one of two random variables is
  stochastically larger than the other. The Annals of Mathematical Statistics
  18(1):50--60

\bibitem[{Mitsch(2023)}]{mitsch2023impact}
Mitsch C (2023) The impact of surplus sharing on the outcomes of specific
  investments under negotiated transfer pricing: An agent-based simulation with
  fuzzy {Q}-learning agents. arXiv preprint arXiv:2301.12255

\bibitem[{Pfeiffer and Wagner(2007)}]{pfeiffer2007rekonstruktion}
Pfeiffer T, Wagner J (2007) {Die Rekonstruktion interner M{\"a}rkte, das
  Dilemma der pretialen Lenkung und spezifische Investitionsprobleme}.
  Schmalenbachs Zeitschrift f{\"u}r Betriebswirtschaftliche Forschung
  59(8):958--981

\bibitem[{Pfeiffer et~al.(2011)Pfeiffer, Schiller, and
  Wagner}]{pfeiffer2011cost}
Pfeiffer T, Schiller U, Wagner J (2011) Cost-based transfer pricing. Review of
  Accounting Studies 16(2):219--246

\bibitem[{Powell and Ryzhov(2012)}]{powell2012optimal}
Powell WB, Ryzhov IO (2012) Optimal learning, vol 841. John Wiley \& Sons

\bibitem[{Rahimiyan and Mashhadi(2006)}]{rahimiyan2006modeling}
Rahimiyan M, Mashhadi HR (2006) Modeling the supplier agent's risk strategy
  based on fuzzy logic combined with the q-learning algorithm. In: 2006
  International Conference on Computational Intelligence and Security, IEEE,
  vol~1, pp 159--163

\bibitem[{Schmalenbach(1908/1909)}]{schmalenbach1908verrechnungspreise}
Schmalenbach E (1908/1909) {{\"U}ber Verrechnungspreise}. Zeitschrift f{\"u}r
  Handelswissenschaftliche Forschung 3(9):165--185

\bibitem[{Simon(1979)}]{simon1979rational}
Simon HA (1979) Rational decision making in business organizations. The
  American Economic Review 69(4):493--513

\bibitem[{Sutton and Barto(2018)}]{sutton2018reinforcement}
Sutton RS, Barto AG (2018) Reinforcement learning: An introduction. MIT Press

\bibitem[{Tijsma et~al.(2016)Tijsma, Drugan, and Wiering}]{tijsma2016comparing}
Tijsma AD, Drugan MM, Wiering MA (2016) Comparing exploration strategies for
  q-learning in random stochastic mazes. In: Symposium Series on Computational
  Intelligence, IEEE, pp 1--8

\bibitem[{Vaysman(1998)}]{vaysman1998model}
Vaysman I (1998) A model of negotiated transfer pricing. Journal of Accounting
  and Economics 25(3):349--384

\bibitem[{Wagner(2008)}]{wagner2008corporate}
Wagner J (2008) {Corporate Governance Strukturen zur L{\"o}sung von
  Koordinations- und Anreizproblemen bei Auftreten spezifischer
  Investitionsprobleme}. PhD thesis, Universit{\"a}t Wien

\bibitem[{Wall(2016)}]{wall2016agent}
Wall F (2016) Agent-based modeling in managerial science: an illustrative
  survey and study. Review of Managerial Science 10(1):135--193

\bibitem[{Welch(1947)}]{welch1947generalization}
Welch BL (1947) The generalization of 'student's' problem when several
  different population variances are involved. Biometrika 34(1/2):28--35

\bibitem[{Wielenberg(2000)}]{wielenberg2000negotiated}
Wielenberg S (2000) Negotiated transfer pricing, specific investment, and
  optimal capacity choice. Review of Accounting Studies 5(3):197--216

\bibitem[{Williamson(1979)}]{williamson1979transaction}
Williamson OE (1979) Transaction-cost economics: the governance of contractual
  relations. The Journal of Law and Economics 22(2):233--261

\bibitem[{Williamson(1985)}]{williamson1985economic}
Williamson OE (1985) The economic institutions of capitalism. New York: Free
  Press

\bibitem[{Young(2001)}]{young2001individual}
Young HP (2001) Individual strategy and social structure: An evolutionary
  theory of institutions. Princeton University Press

\bibitem[{Zadeh(1973)}]{zadeh1973outline}
Zadeh LA (1973) Outline of a new approach to the analysis of complex systems
  and decision processes. IEEE Transactions on systems, Man, and Cybernetics
  (1):28--44

\bibitem[{Zimmermann(2011)}]{zimmermann2011fuzzy}
Zimmermann HJ (2011) Fuzzy set theory and its applications. Springer Science \&
  Business Media

\end{thebibliography}

%
%

\end{document}